\documentclass{article}
\PassOptionsToPackage{numbers, compress}{natbib}

\usepackage[preprint]{neurips_2026}

\usepackage{xspace}
\usepackage{enumitem}
\usepackage{subcaption}
\usepackage{float}      
\usepackage{placeins}   
\newcommand{\methodname}{BenchEvolver\xspace}
\usepackage[utf8]{inputenc} 
\usepackage[T1]{fontenc}    
\usepackage{hyperref}       
\usepackage{url}            
\usepackage{booktabs}       
\usepackage{amsfonts}       
\usepackage{nicefrac}       
\usepackage{microtype}      
\usepackage{xcolor}         
\usepackage{multirow}    
\usepackage{graphicx}    
\usepackage{amsmath,amssymb}
\usepackage{algorithm}
\usepackage{algpseudocode}
\usepackage{longtable}
\usepackage{array}
\usepackage{tabularx}
\usepackage{xurl}
\usepackage[dvipsnames]{xcolor}
\usepackage[most]{tcolorbox}
\usepackage{changepage}
\usepackage{csquotes}   
\tcbuselibrary{listings,breakable,skins}
\usepackage{tikz}
\usepackage{ragged2e}
\usepackage{enumitem}
\usepackage{placeins}
\usepackage{needspace}

\definecolor{PromptBlue}{RGB}{0,110,180}
\definecolor{PromptBack}{RGB}{248,250,252}

\definecolor{SeedBlue}{HTML}{1F3047}
\definecolor{EvoGreen}{HTML}{0C7A5A}
\definecolor{LightGray}{HTML}{F5F7FA}
\definecolor{BorderGray}{HTML}{D9DEE7}

\title{\methodname: Frontier Task Synthesis \\via Solution-Centric Evolution}

\author{%
  Yangzhen Wu\thanks{Equal contribution.}~$^{,1}$ \quad
  Aaron J. Li\footnotemark[1]~$^{,1}$ \quad
  Wenjie Ma$^{1}$ \quad
  Li Cao$^{2}$ \quad
  Ziheng Zhou$^{2}$ \quad \\
  \textbf{Mert Cemri}$^{1}$ \quad
  \textbf{Shu Liu}$^{1}$ \quad
  \textbf{Yuran Xiu}$^{2}$ \quad
  \textbf{Chenxiao Yan}$^{2}$ \quad
  \textbf{Haikun Zhao}$^{2}$ \\
  \textbf{Bin Yu}$^{1}$ \quad
  \textbf{Ion Stoica}\thanks{Equal advising.}~$^{,1}$ \quad
  \textbf{Dawn Song}\footnotemark[2]~$^{,1}$ \\
  $^{1}$University of California, Berkeley \\
  $^{2}$Institute for Interdisciplinary Information Sciences, Tsinghua University \\
  yangzhen\_wu@berkeley.edu \quad\quad aaronjli@berkeley.edu \\
  \vspace{0.4em}\\
  \footnotesize
  \href{https://benchevolver.github.io/}{Project Page}
  \quad\textbar\quad
  \href{https://github.com/thu-wyz/BenchEvolver}{Code}
  \quad\textbar\quad
  \href{https://huggingface.co/BenchEvolver}{Dataset}
}

\begin{document}
\definecolor{NavyBlue}{RGB}{0,80,180}
\newcommand{\problem}{Query2Conf}
\newcommand{\system}{TTPC}

\newcommand{\query}{q}
\newcommand{\config}{\mathsf{c}}
\newcommand{\corr}{y}
\newcommand{\charfn}{\mathcal{F}}
\newcommand{\pcfg}{\hat{p}}
\newcommand{\meancost}{\overline{\text{cost}}}

\newif\ifcomments
\commentstrue
\newcommand{\eat}[1]{} 


\ifcomments
  \providecommand{\shu}[1]{{\color{magenta}{/* shu: #1 */}}}

  \providecommand{\todo}[1]{
    {\colorbox{red}{\bfseries\sffamily\scriptsize\textcolor{white}{TODO}}}
    {\textcolor{red}{\sf\small\textit{#1}}}
  }

  \providecommand{\red}[1]{{\color{red}{/* #1 */}}}
\else
  \providecommand{\shu}[1]{}
  \providecommand{\red}[1]{{\color{red}{/* #1 */}}}
  \providecommand{\todo}[1]{}
\fi


\newcommand{\axisplaceholder}[3][]{%
  \begin{tikzpicture}
    \begin{axis}[
        width=\linewidth,
        height=3.6cm,
        title={#1},
        xlabel={#2},
        ylabel={#3},
        xmin=0, xmax=1,
        ymin=0, ymax=1,
        xtick={0,0.5,1},
        ytick={0,0.5,1},
        grid=both,
        major grid style={line width=0.2pt,draw=gray!25},
        every axis title/.style={font=\small},
        label style={font=\footnotesize},
        tick label style={font=\scriptsize},
        scaled ticks=false,
    ]
      \node[font=\scriptsize\itshape, text=red!70] at (axis cs:0.5,0.5) {placeholder};
    \end{axis}
  \end{tikzpicture}%
}

\newcommand{\axisplaceholderbar}[4][]{%
  \begin{tikzpicture}
    \begin{axis}[
        width=\linewidth,
        height=3.6cm,
        title={#1},
        xlabel={#2},
        ylabel={#3},
        xmin=0, xmax=1,
        ymin=0, ymax=1,
        xtick=\empty,
        ytick={0,0.5,1},
        ymajorgrids,
        major grid style={line width=0.2pt,draw=gray!25},
        every axis title/.style={font=\small},
        label style={font=\footnotesize},
        tick label style={font=\scriptsize},
    ]
      \node[font=\scriptsize\itshape, text=red!70, align=center]
        at (rel axis cs:0.5,0.5) {placeholder\\{\tiny categories: #4}};
    \end{axis}
  \end{tikzpicture}%
}

\maketitle

\begin{abstract}

The rapid progress of frontier large language models has led to widespread benchmark saturation, limiting the ability of existing datasets to differentiate model capabilities or provide useful training signal. For instance, on LiveCodeBench, frontier models achieve over \(99\%\) Pass@1 on easy splits and exceed \(90\%\) Pass@1 on average across difficulty levels~\citep{jain2024livecodebench,gpt55,deepseekai2026deepseekv4}. Constructing new, sufficiently challenging datasets typically requires substantial human effort, creating a bottleneck for continued progress. We introduce \methodname, a solution-centric evolutionary framework that automatically transforms existing coding problems into substantially harder variants. Rather than generating problems from scratch, \methodname evolves reference solutions through structured transformations and derives corresponding problem statements and tests from the evolved solutions. This solution-centric design grounds generation in executable semantics, enabling scalable construction of high-quality, diverse, and difficult tasks with verifiable correctness. Applying \methodname to LiveCodeBench (LCB) and SciCode, we obtain evolved tasks that are substantially more difficult while preserving validity, reference correctness, and diversity. We further curate \textsc{LiveCodeBench-Plus}, a 91-problem benchmark combining evolved and difficult original LCB-v6 tasks, where frontier-model Pass@1 ranges from \(27.5\%\) to \(62.6\%\), restoring clear discrimination among strong coding models. Importantly, evolved tasks remain challenging even for the model that generates them, enabling self-improvement. We further show that RL on evolved LCB tasks improves held-out coding performance: for gpt-oss-20b, seed+evolved training achieves +8.7 and +8.3 Pass@1 gains on LCB v6 Hard and LCB-Pro Easy, exceeding seed-only gains by 70.7\% and 34.8\%, respectively. This closes the loop from self-generated challenges to capability improvement. Our results demonstrate that \methodname can convert saturated benchmarks into frontier-level evaluation suites and reusable training signal.
\end{abstract}

\section{Introduction}
Recent advances in frontier large language models (LLMs) have led to rapid saturation of widely used evaluation benchmarks, limiting their ability to meaningfully measure progress or guide further improvement. On competitive coding benchmarks such as LiveCodeBench~\citep{jain2024livecodebench}, state-of-the-art models now achieve over 99\% pass rate on the newest easy split and exceed 90\% on average across difficulty levels. As a result, these benchmarks provide diminishing discriminative power between models and offer little gradient for training or analysis. This phenomenon is not unique to coding: even when absolute saturation levels differ, static evaluations across reasoning \citep{hendrycks2020measuring,cobbe2021training}, scientific problem solving \citep{rein2023gpqa,tian2024scicode}, and agentic tasks \citep{jimenez2023swe,merrill2026terminal} increasingly lose discriminative power as frontier models improve. Consequently, continued progress increasingly depends on the availability of new, more challenging, and reliably verifiable datasets that co-evolve with frontier models.


Human construction of new benchmarks and datasets is expensive and difficult to scale, creating a bottleneck for continuous model improvement. A natural alternative is synthetic data generation, where LLMs are used to curate new tasks for evaluation and training. Existing methods have made substantial progress in generating synthetic questions from seed data \citep{luo2023wizardcoder,xu2025wizardlm,wei2024magicoder,wei2024selfcodealign,ahmad2025opencodereasoning,pham2025swe,sancaktar2026deep}. However, many of these pipelines follow an asymmetric teacher--student paradigm: a strong model synthesizes, filters, or verifies data that is then used to train or evaluate weaker models. In addition, much of the synthesis operates at the instruction level, improving prompt diversity and surface complexity without necessarily changing the underlying solution structure or providing explicit control over task difficulty. Moving from synthetic instructions to reusable benchmarks therefore requires generating complete executable tasks, where the statement, reference solution, and tests are jointly valid. Recent work has begun to address this setting, but correctness and test validity are often ensured through stronger models, external validation, or human verification~\citep{zhou2025autocode}. Consequently, existing pipelines do not directly address the self-challenging setting required for open-ended improvement: generating tasks that are valid, verifiable, difficulty-controlled, and hard for the generator itself. Without this property, synthetic data generation remains primarily a way to distill stronger models into weaker ones, rather than a mechanism by which frontier models can expose their own weaknesses and improve through training on self-generated challenges.


Toward this end, we propose \methodname, a solution-centric evolutionary framework for transforming saturated coding tasks into harder yet verifiable problems. Rather than generating a new problem statement first, \methodname evolves the reference solution itself, using the evolved solution as an executable oracle from which the statement, examples, and tests are derived. Each accepted mutation must change the solution structure enough to make the parent algorithm insufficient, and candidates are accepted only after independent consistency checks and empirical difficulty evaluation against a target model panel. This design measures task difficulty empirically rather than heuristically, and enables frontier models to generate challenges that expose their own weaknesses without relying on a strictly stronger teacher.


We apply our framework to LiveCodeBench (LCB)~\citep{jain2024livecodebench}, a competitive-programming benchmark, and SciCode~\citep{tian2024scicode}, a research-oriented scientific coding benchmark. Across both domains, \methodname generates valid evolved tasks at scale and substantially reduces target-model pass rates. We also construct \textsc{LiveCodeBench-Plus}, a difficulty-upgraded coding benchmark that combines validated evolved tasks with challenging original LiveCodeBench problems to provide a harder and more discriminative evaluation set. We further show that the solution-centric design outperforms a problem-centric generation baseline, that memory-guided evolution improves over independent one-step mutations, and that the evolved tasks can support closed-loop self-improvement through reinforcement learning. In particular, using \texttt{gpt-oss-20b} as both the evolver and the target model, we find that training on evolved problems improves held-out coding performance more than training on the original seed problems alone, and that the combined seed+evolved mixture performs best. Across the two informative held-out evaluation settings we study, the combined mixture yields a \(70.7\%\) and \(34.8\%\) larger improvement over the seed-only RL baseline, respectively. These results suggest that evolved tasks are not merely harder evaluation items, but can also provide useful training signal for improving the same model family that generated them.

Our contributions are summarized as follows:
\begin{itemize}
    \item We introduce \methodname, a solution-centric framework that upgrades saturated coding tasks into harder variants grounded in executable reference solutions and verifiable tests.

    \item We show that \methodname works across competitive-programming and scientific-coding domains, producing valid, diverse, and substantially harder tasks on LiveCodeBench and SciCode; we further use it to construct \textsc{LiveCodeBench-Plus}, a difficulty-upgraded benchmark that better discriminates among frontier models.

    \item We provide initial evidence that self-challenging task evolution can support model improvement. For \texttt{gpt-oss-20b}, training on seed+evolved tasks or evolved tasks alone improves held-out coding performance more than training on the original seed set, indicating that evolved tasks can serve as reusable RL signal rather than only as harder benchmark items.
\end{itemize}
\section{Related Work}
Our work connects synthetic coding data, self-play, and evolutionary search; we provide a more comprehensive related work discussion in Appendix~\ref{appendix:additional_related_work}. Prior work synthesizes code instructions, reasoning traces, repository-level bug-fix tasks, competitive-programming problems, and RL curricula to improve code models~\citep{luo2023wizardcoder,xu2025wizardlm,wei2024selfcodealign,ahmad2025opencodereasoning,pham2025swe,sancaktar2026deep,zhou2025autocode,wu2026xcoder}. A complementary line of work uses model-in-the-loop data generation for self-improvement, including self-challenging tool-use agents, solver--conjecturer self-play, code-centric data synthesis, and adversarial co-evolution between code and test generators~\citep{zhou2025self,bailey2026scaling,sun2025codeevo,zhang2026embarrassingly,wang2026codea1}. Finally, LLM-based evolutionary methods have been used for prompt optimization, program discovery, algorithm search, and self-evolving agents~\citep{fernando2023promptbreeder,guo2023evoprompt,agrawal2025gepa,romera2024funsearch,novikov2025alphaevolve,hu2024automated,wu2025evolveR}. \methodname differs in the object being evolved: rather than generating auxiliary supervision or optimizing solutions for a fixed task, it evolves complete executable benchmark items---statements, reference solutions, and tests---selected by empirical target-model failure. This turns inference-time search~\citep{wu2024inference} into reusable benchmark and training data for closed-loop self-improvement.
\section{Method}
\label{sec:method}

\begin{figure}[t]
    \centering
    \includegraphics[width=\linewidth,
    trim={0.5cm 0cm 0.8cm 0cm},
    clip]{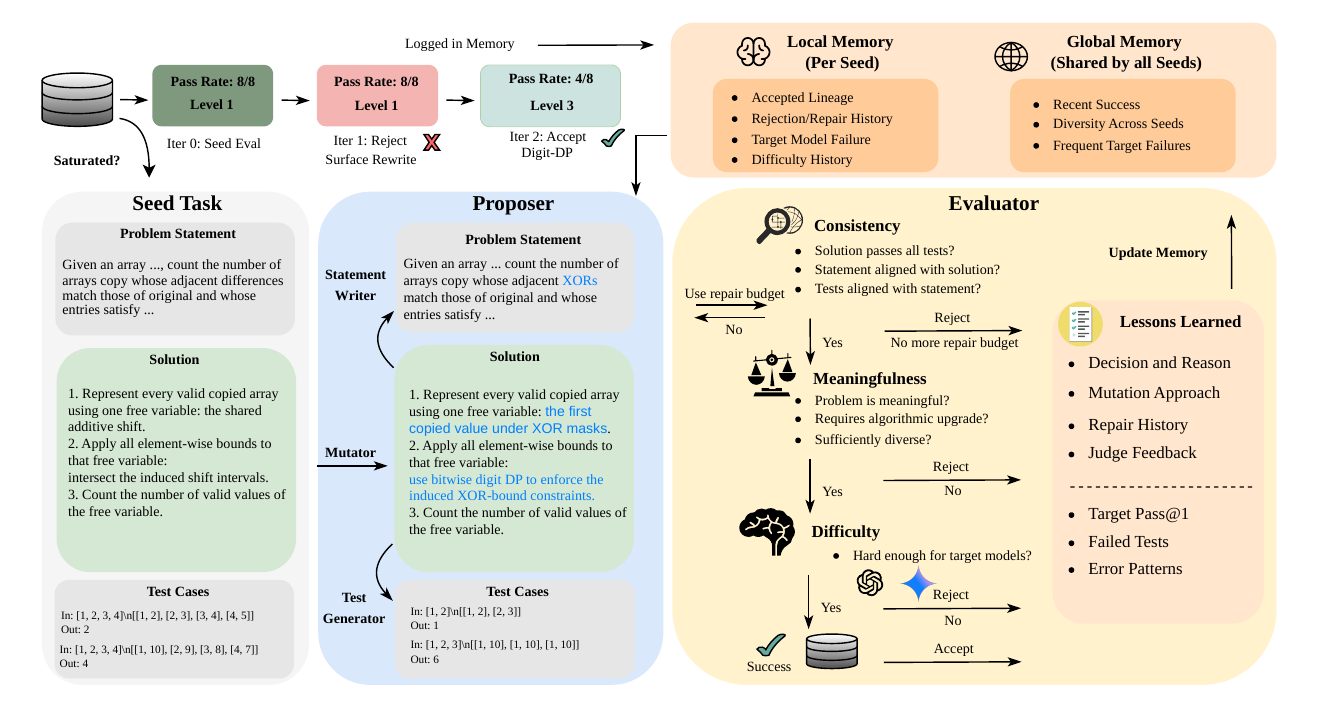}
    \caption{Overview of \methodname. Starting from a saturated seed task, the proposer first mutates the reference solution and derives a new statement and tests; then the evaluator filters candidates for validity, diversity, and difficulty; memory is updated to include evolution outcomes with reasons, and accepted candidates become new parents.}
    \label{fig:framework}
\end{figure}

We introduce \methodname, a closed-loop framework for evolving saturated programming benchmarks into harder, verifiable tasks. The framework follows three principles: \emph{generate in solution space}, \emph{verify by independent consistency checks}, and \emph{select by empirical model failure}. Given seed tasks, a \emph{Proposer} constructs candidate evolutions, an \emph{Evaluator} validates them and measures target-model difficulty, and a \emph{Memory} module feeds past successes and failures back into search. Figure~\ref{fig:framework} provides an overview.

\subsection{Self-Challenging Problem Evolution}
\label{subsec:self_challenging_setting}

We consider a benchmark \(\mathcal{D}=\{I_i\}\) of executable programming tasks. Each task is represented as
\[
    I = (S, C, T, E),
\]
where \(S\) is the natural-language statement, \(C\) is a reference implementation, \(T\) is a hidden test suite, and \(E\) is an execution harness. The harness is the only domain-specific component: for competitive-programming tasks, \(E\) runs code on stdin and checks stdout; for scientific coding tasks, \(E\) executes function-level submissions against assertion tests and domain-specific oracle artifacts when available. The rest of the framework---mutation, writing, verification, difficulty measurement, and memory-guided search---is shared across domains.

Our goal is to construct an evolved benchmark \(\mathcal{D}'\) whose tasks are well-posed, verifiable, empirically harder than their seeds, and topically and algorithmically diverse. Crucially, this construction should not require a model stronger than the target panel; otherwise, the setting reduces to teacher--student distillation rather than self-challenging generation.

We define difficulty behaviorally: each target model receives multiple attempts, and an attempt succeeds only if the generated program passes all hidden tests. The average success rate across models and attempts is the empirical pass rate, with lower pass rate indicating higher difficulty. Thus, hardness is measured by executable model failure rather than assigned by the generator or an LLM judge.

\subsection{Proposer: Solution-Centric Task Generation}
\label{subsec:proposer}

The Proposer constructs candidate tasks by evolving solutions rather than statements. Conventional synthetic problem generation often begins with a new statement, after which the system must infer or verify whether a correct solution and tests exist. This statement-first direction is fragile in the self-challenging setting: when the same model proposes and verifies a task, ambiguous specifications, hidden assumptions, and superficial complexity can pass undetected. \methodname instead follows a solution-first pipeline:
\[
    C \;\longrightarrow\; C' \;\longrightarrow\; I'=(S', C', T', E),
\]
where the parent reference solution \(C\) is first mutated into an evolved solution \(C'\), and the statement \(S'\) and tests \(T'\) are then derived around this evolved solution under the benchmark's fixed execution harness \(E\).

This design makes difficulty a property of the task's computation rather than its surface form. The Proposer is instructed to introduce a \emph{dominant algorithmic lift}: a substantive change to the solution structure that makes the parent approach insufficient. Intuitively, such lift often arises when the evolved task requires a stronger asymptotic strategy, a richer data structure or state-maintenance mechanism, a new structural or mathematical reformulation, or natural constraints that invalidate the parent task's simple shortcut. These intuitions guide solution-space mutation toward meaningful computational changes, while accepted tasks are still judged by executable consistency and empirical target-model failure rather than by surface complexity, longer statements, adversarial formats, or obscure edge cases.

Each proposal is produced by first mutating the parent reference implementation into an evolved solution \(C'\), together with a concise explanation of the new algorithmic idea and why the parent solution fails. The Proposer then derives a natural-language statement \(S'\), public examples, and hidden tests around \(C'\). Public examples and expected test outputs are materialized by executing the evolved reference solution, anchoring the task in executable behavior. For stdin--stdout tasks, tests are organized into small, medium, large, and stress regimes; for scientific function-level tasks, they are assertion-style tests executed by the domain harness.

The Proposer is conditioned on feedback from previous attempts, including parent difficulty, accepted mutations, rejection reasons, target-model error patterns, and global diversity signals. It does not decide whether a candidate is valid or difficult; its role is to propose a complete executable task, while acceptance is left to the Evaluator.

\subsection{Evaluator: Verification and Empirical Selection}
\label{subsec:evaluator}

The Evaluator ensures that a candidate is both valid and genuinely challenging. Validity means that the statement, reference solution, tests, and execution harness define the same task; difficulty is measured only after this consistency is established.

For validation, \methodname uses benchmark-specific checks rather than relying on a single judge. In competitive programming, it triangulates among the evolved reference solution, a statement-only brute-force solver, and a statement-only public-output oracle. Since these witnesses observe different information, their disagreements help identify whether the issue lies in the reference, the brute-force solver, the public outputs, or the specification. For scientific coding, where brute-force solvers are less applicable, \methodname uses a statement-faithfulness check: an independently generated solution from the statement is run against the candidate tests to assess whether the written task sufficiently determines the intended computation. Full validation and repair details are given in Appendix~\ref{app:validation}.

After validation, each target model receives multiple attempts, and an attempt succeeds only if it passes all hidden tests. The resulting pass rate is mapped to the same difficulty scale used for seed tasks. A candidate is accepted only if it improves over the seed difficulty, and optionally over the parent difficulty, making selection empirical rather than judge-assigned.

Finally, the Evaluator filters out false difficulty: ambiguous wording, misleading I/O, underspecified constraints, unnatural edge cases, or near-duplicate reskins. Localized failures are routed to bounded repair; candidates that cannot be made consistent within the repair budget are rejected. Thus, accepted tasks must be both executable and empirically hard.

\subsection{Memory-Guided Evolution}
\label{subsec:memory}

Memory turns \methodname from repeated sampling into adaptive search. Each seed maintains a local memory of its lineage, including accepted mutations, failed attempts, validation issues, target-model pass rates, and observed error patterns. This history is summarized and fed back to the Proposer, helping later mutations avoid repeated failures and focus on directions that have exposed model weaknesses.

\methodname also maintains a global memory across seeds. This memory records accepted and attempted mutation families throughout the run, encouraging different lineages to explore distinct algorithmic directions rather than rediscovering the same lift under different surface forms. It is also used in selection: when a mutation family has already succeeded elsewhere, a new candidate from that family must provide a larger difficulty gain to be accepted. Diversity is therefore enforced both through generation context and through the acceptance rule.

Together, local and global memory give \methodname a lightweight evolutionary structure. Mutation comes from solution-centric proposal; selection comes from validation and empirical target-model failure; inheritance comes from accepted lineages; and diversity is maintained through memory. The resulting tasks form a hard, verified distribution that can serve both as an evolved benchmark and as reinforcement-learning data, enabling a closed loop in which models generate challenges, train on executable rewards, and return to generate harder tasks.

The full pseudocode of \methodname is provided in Appendix~\ref{appendix:pseudocode}.

\section{Experiments}
\label{sec:experiments}

We evaluate \methodname along three dimensions. First, we study whether it can generate valid, diverse, and empirically harder tasks across two executable coding domains: competitive programming and scientific coding. Second, we describe \textsc{LiveCodeBench-Plus}, a benchmark artifact constructed to provide challenging evaluation for frontier coding models. Third, we test whether evolved tasks provide useful reinforcement-learning signal for improving the same model family that generates them.

\subsection{Task-Evolution Evaluation across Executable Coding Domains}
\label{subsec:task_evolution_results}

For task-evolution evaluation, we consider two target tiers. The \emph{lightweight} tier consists of GPT-5.4-mini and Gemini-3-Flash, while the \emph{frontier} tier consists of GPT-5.4 and Gemini-3.1-Pro. We exclude Claude models from the target pool because they perform slightly worse on both benchmarks; including them could cause evolution to target Claude-specific failure modes rather than difficulty representative of the intended target tier. Details of the evolution configurations and hyperparameter choices are provided in Appendix ~\ref{app:evolution_config}.

\subsubsection{Competitive programming: LiveCodeBench}
For the controlled LiveCodeBench comparison, we randomly sample 65 seed problems spanning easy, medium, and hard difficulty levels from v6, balancing difficulty coverage against the substantial cost of multi-model evolution and target evaluation. We use GPT-5.4-mini, Gemini-3-Flash, and Claude-Sonnet-4.6 as evolvers in the lightweight-target setting. To test whether \methodname can generate tasks that remain challenging for frontier targets, we additionally select 10 saturated problems from each difficulty level and use Gemini-3.1-Pro as the evolver. Before target-model evaluation, all candidates are validated using the LiveCodeBench-specific brute-force triangulation protocol described in Appendix~\ref{app:lcb_validation}.

\begin{table*}[t]
\centering
\scriptsize
\setlength{\tabcolsep}{4.5pt}
\renewcommand{\arraystretch}{1.08}
\begin{tabular}{lllcccc}
\toprule
Method & Target Tier & Evolver Model
& \multicolumn{3}{c}{Evolved Seed Fraction}
& Validity \\
\cmidrule(lr){4-6}
& & & Easy & Medium & Hard & Avg. \\
\midrule
Problem-Centric & Lightweight & GPT-5.4-mini
& 12/22 (54.5\%) & 11/23 (47.8\%) & 6/20 (30.0\%)
& 79.3\% \\

Memory-Free & Lightweight & GPT-5.4-mini
& 13/22 (59.1\%) & 15/23 (65.2\%) & 9/20 (45.0\%)
& 86.2\% \\
\midrule
\methodname & Lightweight & GPT-5.4-mini
& 17/22 (77.3\%) & 16/23 (69.6\%) & 12/20 (60.0\%)
& 97.7\% \\

\methodname & Lightweight & Gemini-3-Flash
& 18/22 (81.9\%) & 14/23 (60.9\%) & 16/20 (80.0\%)
& 89.9\% \\

\methodname & Lightweight & Claude-Sonnet-4.6
& 14/22 (63.6\%) & 15/23 (65.2\%) & 7/20 (35.0\%)
& 93.4\% \\

\methodname & Frontier & Gemini-3.1-Pro
& 8/10 (80.0\%) & 9/10 (90.0\%) & 6/10 (60.0\%)
& 96.7\% \\
\bottomrule
\end{tabular}
\caption{
LiveCodeBench-v6 evolution yield and validity by seed difficulty. Easy, Medium, and Hard report the fraction of completed seeds for which at least one accepted evolved problem is generated. Validity is the mean post-hoc validity rate of accepted evolved problems judged by Claude Code Opus 4.7. Problem-Centric and Memory-Free are ablations using GPT-5.4-mini as the evolver.
}
\label{tab:lcb-yield}
\end{table*}

\begin{figure}[t]
    \centering
    \includegraphics[width=\linewidth]{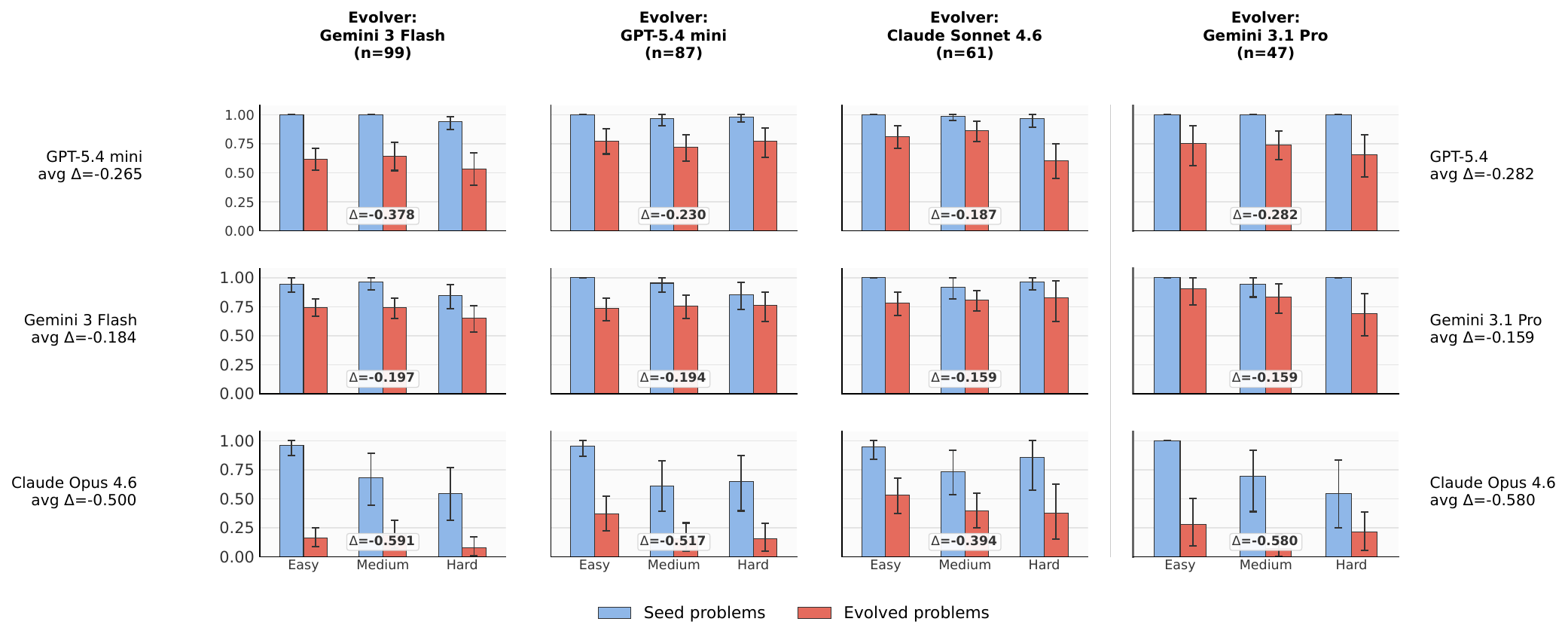}
    \caption{Pass@1 on original LiveCodeBench seed problems versus evolved problems. Each column group corresponds to an evolver model, and each row corresponds to a target model. Within each subfigure, bars show Easy, Medium, and Hard seeds from left to right. Evolved problems consistently reduce target-model pass rates across both lightweight and frontier models (\(k=4\) attempts per model).}
    \label{fig:lcb_diff}
\end{figure}

\paragraph{Synthesis yield, validity, and ablations.}
Table~\ref{tab:lcb-yield} shows that \methodname reliably transforms saturated LiveCodeBench seeds into valid evolved tasks across difficulty levels and evolver models. Under the same GPT-5.4-mini evolver, the full method substantially improves both evolved-seed coverage and post-hoc validity over the problem-centric baseline. This supports the solution-centric design: evolving executable algorithmic logic before recovering a statement makes it easier to construct coherent problems and tests than generating statements first and validating them afterward. The memory-free ablation also underperforms \methodname, indicating that accepted lineages and prior failures provide useful search guidance beyond independent one-step mutations.

\paragraph{Evolved tasks are empirically harder.}
Figure~\ref{fig:lcb_diff} shows that accepted evolved tasks substantially reduce target-model pass rates relative to their original seeds across both lightweight and frontier models. The effect is consistent across difficulty levels and evolver models, indicating that the tasks are not artifacts of a single generator or prompt configuration. Crucially, each evolver also experiences a clear accuracy drop on its own evolved tasks. Thus, \methodname does not merely use a stronger model to generate data for weaker models; it constructs verified tasks that expose weaknesses of the model generating them.

\begin{figure}[t]
\centering
\includegraphics[width=0.85\linewidth]{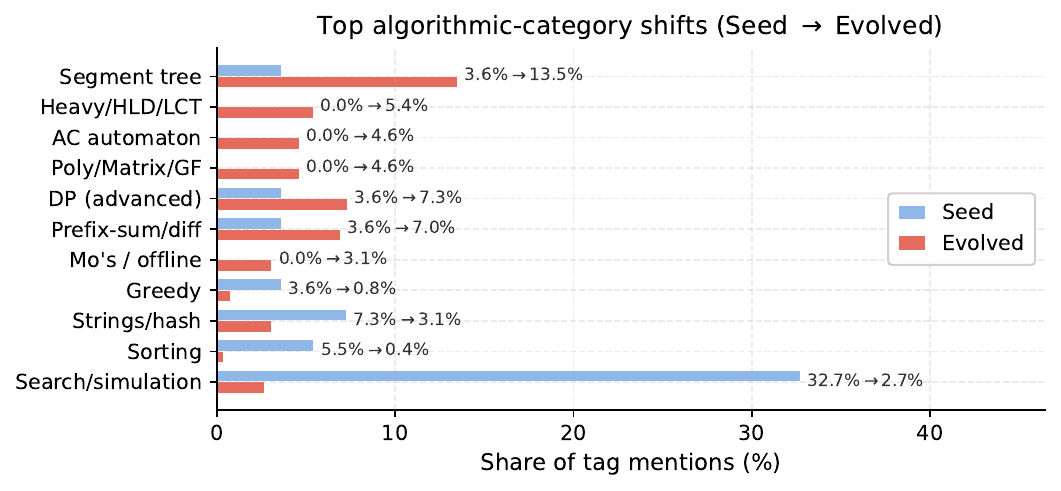}
\caption{Top eleven algorithm/data-structure categories ordered by absolute
seed$\to$evolved share shift. Numbers next to each bar pair show seed-share$\,\to\,$evolved-share. Full category breakdown is in Appendix~\ref{app:human_eval_full}.}
\label{fig:human_eval_tags}
\end{figure}

\paragraph{Human evaluation: algorithmic diversity.}
\label{paragraph:human_eval}
We complement executable evaluation with a blind human study of LiveCodeBench evolutions produced in our broader method-evaluation pool, which includes additional lineages beyond the 65-seed controlled comparison above. Six competitive-programming experts (Codeforces master / IOI / ICPC level) reviewed 100 evolved seed lineages spanning \(72\) distinct LiveCodeBench seeds and \(207\) distinct evolved problems, identifying the algorithms and data structures required by each task. Full protocol details and additional ratings of clarity, novelty, difficulty, and validity are provided in Appendix~\ref{app:human_eval_full}.

As shown in Figure~\ref{fig:human_eval_tags}, the seed problems are dominated by a single algorithmic regime: \emph{Search/simulation} accounts for \(32.7\%\) of seed-tag mentions. In contrast, evolved problems distribute mass across a broader range of advanced data structures and algorithmic regimes, including HLD/LCT, AC automata, and polynomial/matrix methods. The number of distinct algorithmic categories increases from \(19\) in the seeds to \(30\) in the evolved set; moreover, \(95.6\%\) of reviewed lineages introduce at least one category absent from their seed, with \(2.54\) new categories per lineage on average. These results show that \methodname does not merely increase difficulty through superficial modification: it broadens the algorithmic surface area on which target models are challenged.

\subsubsection{Scientific coding: SciCode}
We next evaluate whether the solution-centric principle extends beyond competition-style stdin--stdout programs. For SciCode, we select 30 self-contained subproblems spanning 15 main problems from the validation split. Among them, 27 are saturated by the lightweight target tier and 28 by the frontier target tier, serving as seeds for generating harder scientific-coding variants. Since SciCode does not naturally admit brute-force validation, we instead use the customized statement-faithfulness protocol described in Appendix~\ref{app:scicode_validation}, together with assertion-based execution.

\begin{table*}[t]
\centering
\scriptsize
\begin{tabular}{llcc}
\toprule
Target Tier & Evolver Model & Evolved Seed Fraction & Validity \\
\midrule
Lightweight & GPT-5.4-mini       & 24/27 (88.9\%) & 91.0\% \\
Lightweight & Gemini-3-Flash     & 26/27 (96.3\%) & 98.2\% \\
Lightweight & Claude-Sonnet-4.6 &  8/27 (29.6\%) & 90.0\% \\
Frontier    & Gemini-3.1-Pro     & 17/28 (60.7\%) & 88.9\% \\
\bottomrule
\end{tabular}
\caption{
SciCode evolution yield and validity. Evolved Seed Fraction reports the fraction of saturated seeds for which at least one accepted evolved problem is generated; Validity reports the post-hoc validity rate of accepted evolved problems.
}
\label{tab:scicode-yield}
\end{table*}

\begin{figure}[t]
    \centering
    \includegraphics[width=\linewidth]{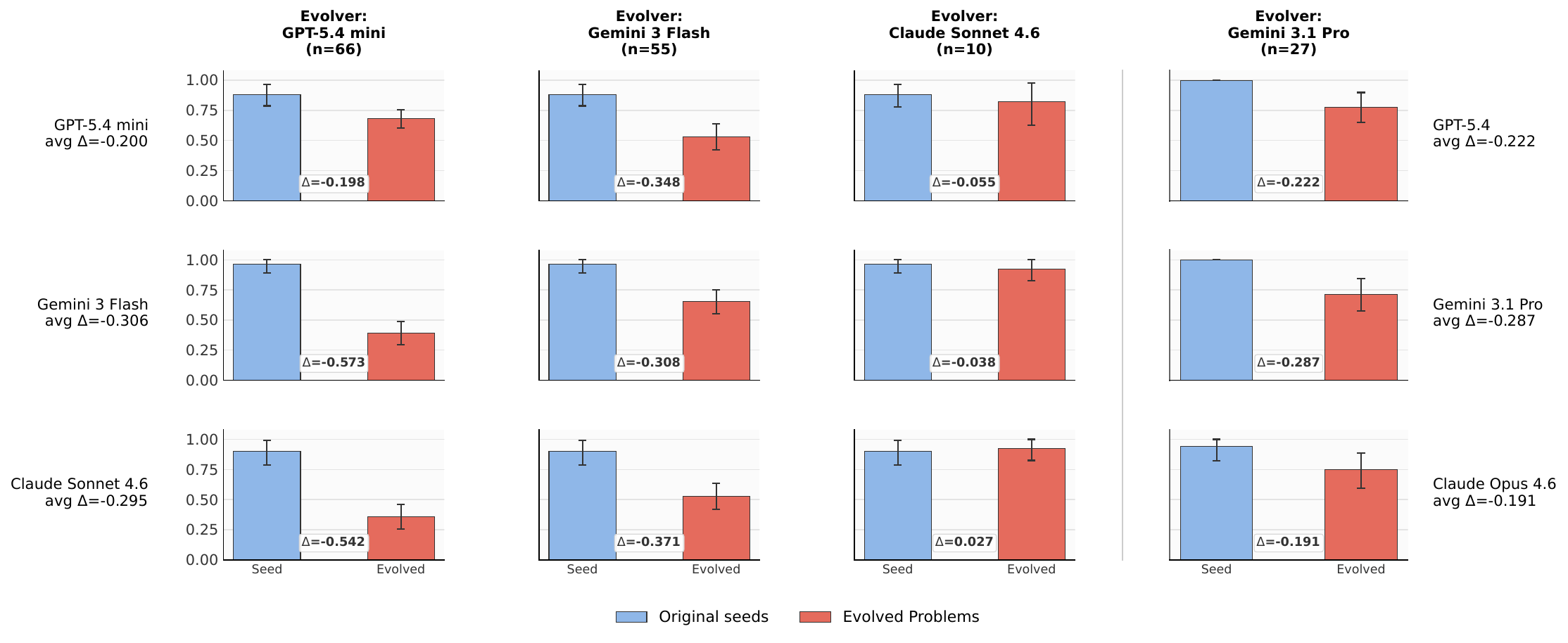}
    \caption{Pass@1 on original SciCode seed problems versus evolved problems. Across both lightweight and frontier target models, evolved problems substantially reduce model pass rates (\(k=4\) attempts per model).}
    \label{fig:scicode_diff}
\end{figure}

Table~\ref{tab:scicode-yield} and Figure~\ref{fig:scicode_diff} show that \methodname also generates valid and substantially harder scientific-coding tasks. Despite the smaller seed pool and the absence of brute-force oracles, the framework produces accepted evolved tasks with high validity across evolvers while consistently reducing target-model pass rates relative to the original seeds. This indicates that solution-centric evolution is not specific to competitive programming: the same generation principle extends across executable coding domains when paired with a validation protocol appropriate to the benchmark harness.

\subsection{\textsc{LiveCodeBench-Plus}: A Benchmark for Frontier Coding Models}
\label{subsec:lcb_evolved_benchmark}

In this section, we apply \methodname to saturated problems from the medium and hard tasks in LiveCodeBench-v6, in order to construct a difficulty-upgraded benchmark useful for frontier model evaluation, which we call \textsc{LiveCodeBench-Plus}.

\paragraph{Evolution and evaluation.}
We start by reusing the same seed saturation criteria in Section \ref{subsec:task_evolution_results}. Of the 52 medium problems evaluated against the lightweight tier, 31 are saturated; of these, 23 produce at least one accepted evolved problem. Of 80 v6-hard problems evaluated against the frontier tier, 57 are saturated; of these, 43 produce at least one accepted evolved problem. We evolve the medium split using Gemini-3-Flash and the Hard split using Gemini-3.1-Pro, while using the same \textit{lightweight} and \textit{frontier} target tiers from Section \ref{subsec:task_evolution_results}. We keep all other evolution configurations the same in Appendix~\ref{app:evolution_config}. To ensure our benchmark remains challenging for the strongest frontier models, we expand the evaluation pool to eight models across multiple providers.

\paragraph{Validation and Filtering.}
In addition to the internal brute-force triangulation protocol described in Appendix~\ref{app:lcb_validation}, each evolved problem is reviewed by human evaluators on correctness and meaningfulness. We only select problems that pass all of the following criteria to be included in \textsc{LiveCodeBench-Plus}: 
\begin{itemize}
    \item \textbf{Quality gate}: Quality gate: the problem receives a comprehensive quality and novelty score of at least 3 on a 1–5 scale, following coding-olympiad standards, ensuring the evolved problem is also high-quality for human coding competitors.
    \item \textbf{Difficulty range}: the combined model \textit{pass@1} across our evaluation suite is restricted from $0.05$ to $0.75$, excluding problems that are potentially degenerate and meaningless, or are too easy to discriminate among strong models.
\end{itemize}
After filtering, we retain 64 evolved problems (44 from the hard split, 20 from the Medium split). We supplement these with 27 problems drawn from the original LiveCodeBench-v6 subset that satisfy the same difficulty ceiling ($\leq 0.75$), yielding a final benchmark of 91 problems.

\paragraph{Resulting benchmark.}
In total, our evolution pipeline produces LiveCodeBench Evolved: 35 validated evolved tasks from 23 Medium seeds and 55 from 43 Hard seeds, all preserving the executable stdin--stdout interface of LiveCodeBench. After applying the quality and difficulty filters described above, we retain 20 Medium and 44 Hard evolved problems; combined with 27 difficult original LiveCodeBench-v6 problems, this yields \textsc{LiveCodeBench-Plus}, a benchmark of 91 problems spanning a wide range of advanced algorithmic topics. The 35-problem LiveCodeBench Evolved Medium is also used as an external held-out evaluation set in Section~\ref{subsec:self_improvement_rl}. We release the benchmarks and all problems in our Hugging Face repository.

\paragraph{Difficulty shift.}
Table~\ref{tab:lcb-plus} shows that evolution substantially increases difficulty relative to the source seeds. On the Hard split, average \textit{pass@1} across all evaluated models drops from $87.0\%$ on the source seeds to $45.7\%$ on the evolved tasks, an absolute reduction of $41.3$ points. The Medium split shows a consistent but smaller shift, from $96.5\%$ to $69.6\%$, a $26.8$-point reduction. This decrease holds for every individual model: for example, GPT-5.4 drops from $94.8\%$ to $49.7\%$ on the Hard split, while DeepSeek-V4-Pro drops from $83.7\%$ to $23.2\%$. On the full 91-problem \textsc{LiveCodeBench-Plus} benchmark, pass@1 ranges from $27.5\%$ (DeepSeek-V4-Pro) to $62.6\%$ (GPT-5.5), confirming that the combined set remains challenging even for the strongest frontier models and provides clear discrimination across the evaluated models.

\begin{table*}[t]
    \centering
    \small
    \setlength{\tabcolsep}{5pt}
    \renewcommand{\arraystretch}{1.08}
    \begin{tabular}{lccccccc}
    \toprule
     & \multicolumn{3}{c}{\textbf{Medium} (23 seeds / 35 evolved)}
     & \multicolumn{3}{c}{\textbf{Hard} (43 seeds / 55 evolved)}
     & \textbf{LCB-Plus} \\
    \cmidrule(lr){2-4}\cmidrule(lr){5-7}
    \textbf{Model}
     & Seed & Evolved & $\Delta$
     & Seed & Evolved & $\Delta$
     & (91 problems) \\
    \midrule
    GPT-5.5          & 100.0 & 80.0 & -20.0 & 97.1 & 62.3 & -34.8 & 62.6 \\
    GPT-5.4          & 98.9  & 74.3 & -24.6 & 94.8 & 49.7 & -45.1 & 54.1 \\
    GPT-5.4-mini     & 95.7  & 59.3 & -36.4 & 79.7 & 21.7 & -58.0 & 29.6 \\
    Gemini-3.1-Pro   & 100.0 & 78.6 & -21.4 & 96.5 & 56.8 & -39.7 & 59.1 \\
     Gemini-3.5-Flash & 95.7  & 73.6 & -22.1 & 87.8 & 50.5 & -37.3 & 50.0 \\
    Gemini-3-Flash   & 88.0  & 67.1 & -20.9 & 82.0 & 43.2 & -38.8 & 40.1 \\
    DeepSeek-V4-Pro  & 95.7  & 57.1 & -38.6 & 83.7 & 23.2 & -60.5 & 27.5 \\
    Qwen-3.7-Max     & 97.8  & 67.1 & -30.7 & 74.4 & 58.6 & -15.8 & 47.5 \\
    \bottomrule
    \end{tabular}
    \caption{Pass@1 (\%) on seed problems and their evolved variants across medium and hard
    difficulty tiers ($k{=}4$ attempts per model), and on \textsc{LiveCodeBench-Plus}
    (91 problems combining evolved tasks and original hard LCB-v6 problems).
    \emph{Seed} is the macro-average pass@1 over the original LiveCodeBench-v6 problems
    from which each evolved set derives; \emph{Evolved} is the macro-average over the
    corresponding evolved problems. $\Delta$ is the absolute drop. For cost and latency reasons, we evaluate GPT models with medium reasoning effort,
Gemini models with adaptive reasoning, and DeepSeek-V4-Pro with high thinking mode; all API calls use a 600-second timeout.}
    \label{tab:lcb-plus}
  \end{table*}

\subsection{Self-Improvement through Reinforcement Learning}
\label{subsec:self_improvement_rl}

The results above establish that \methodname can produce verified tasks that challenge current models. We next ask whether evolved tasks can also expose weaknesses that are useful for improving the same model through training. This is the central promise of self-challenging data generation: rather than relying on a stronger teacher, a model constructs executable challenges near its own capability boundary and then learns from the resulting reward signal.

\paragraph{Setup.}
To test this, we use \texttt{gpt-oss-20b} as both the evolver and the target model. We take 880 LiveCodeBench v1--v5 problems released before January 2025 as the seed pool and hold out LiveCodeBench v6 and LiveCodeBench-Pro as evaluation sets. We apply \methodname only to seeds that \texttt{gpt-oss-20b} solves correctly in all five initial attempts, i.e., problems saturated for the model. For each eligible seed, we allow up to 10 evolution iterations and accept at most two evolved problems; an accepted task must reduce the model's empirical accuracy while remaining nonzero, ensuring that it is challenging but not degenerate. This procedure yields 586 evolved problems from 404 successfully evolved seeds.

We construct three RL training sets: the original seed set with 880 problems, the evolved set with 586 problems, and their union with 1,466 problems. We train \texttt{gpt-oss-20b} with GRPO using Tinker, running two independent random seeds for each data condition under identical configurations: 64 problems per batch, 16 rollouts per problem, and maximum output lengths of 24K tokens during training and 30K tokens during evaluation. We report held-out performance on LiveCodeBench v6 Hard (80 problems) and LiveCodeBench-Pro Easy (96 problems), which lie in the informative difficulty range for \texttt{gpt-oss-20b}; other splits are either nearly saturated or too difficult to support meaningful comparisons. In addition, we evaluate on \textsc{LCB-Evolved} Medium, an independently constructed evolved evaluation set generated with Gemini-3-Flash rather than \texttt{gpt-oss-20b}. This split is not used in RL training and has nonzero but far-from-saturated base-model accuracy, making it an informative external test of whether self-generated training signal transfers to harder evolved tasks.

\begin{figure*}[t]
    \centering
    \begin{subfigure}[t]{0.32\linewidth}
      \centering
      \includegraphics[width=\linewidth]{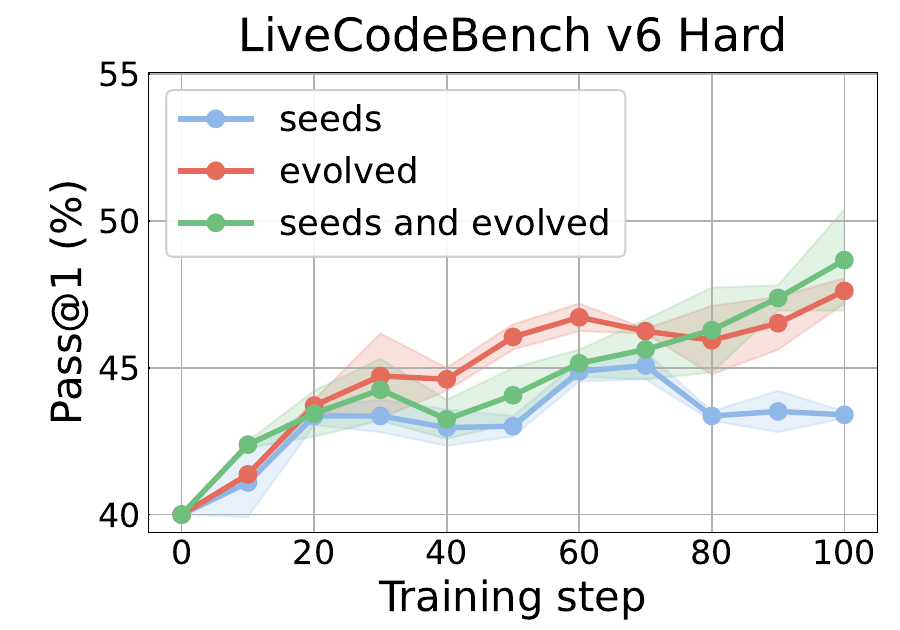}
      \caption{LCB v6 Hard}
      \label{fig:lcb_v6_hard}
    \end{subfigure}
    \hfill
    \begin{subfigure}[t]{0.32\linewidth}
      \centering
      \includegraphics[width=\linewidth]{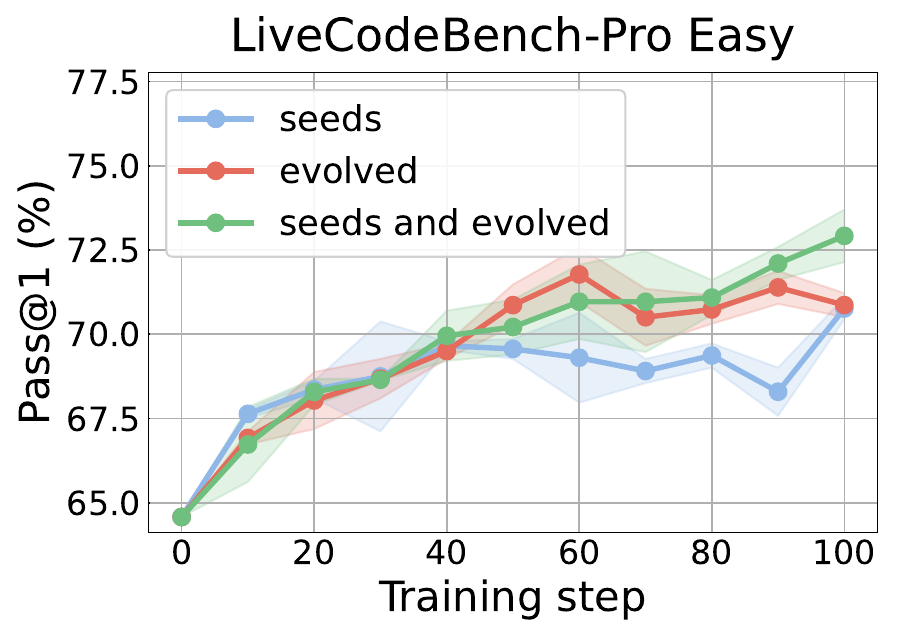}
      \caption{LCB-Pro Easy}
      \label{fig:lcb_pro_easy}
    \end{subfigure}
    \hfill
    \begin{subfigure}[t]{0.32\linewidth}
      \centering
      \includegraphics[width=\linewidth]{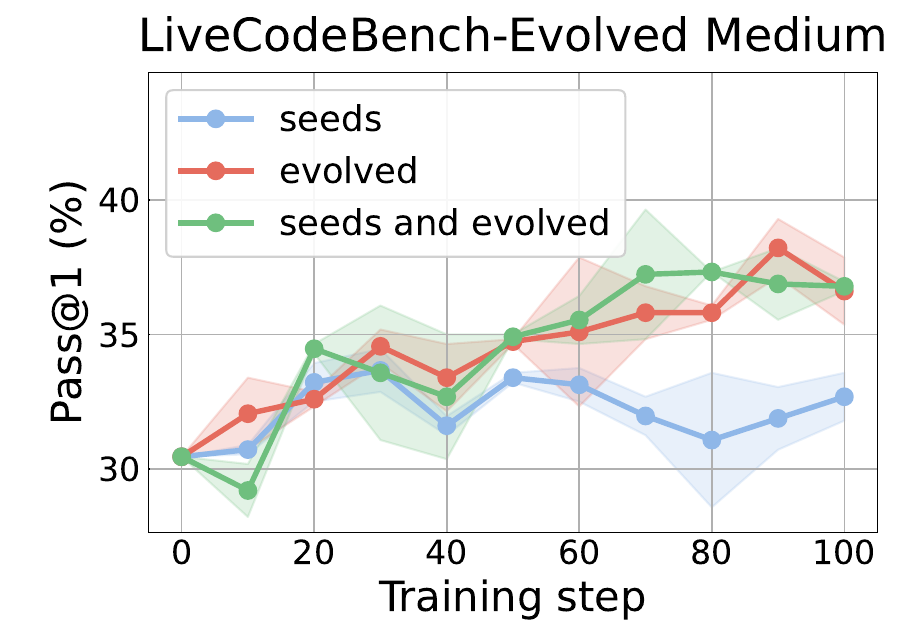}
      \caption{\textsc{LCB-Evolved} Medium}
      \label{fig:lcb_evolved_medium}
    \end{subfigure}
    \caption{Test accuracy across training steps for three RL data mixes (mean \(\pm\) standard deviation over two random seeds). Step 0 shows the base model. Pass@1 is computed by averaging 16 samples per problem. \textsc{LCB-Evolved} Medium is constructed independently using Gemini-3-Flash as the evolver and is not used in RL training.}
    \label{fig:acc_curves}
\end{figure*}

\begin{figure*}[t]
      \centering
      \begin{subfigure}[t]{0.32\linewidth}
        \centering
        \includegraphics[width=\linewidth]{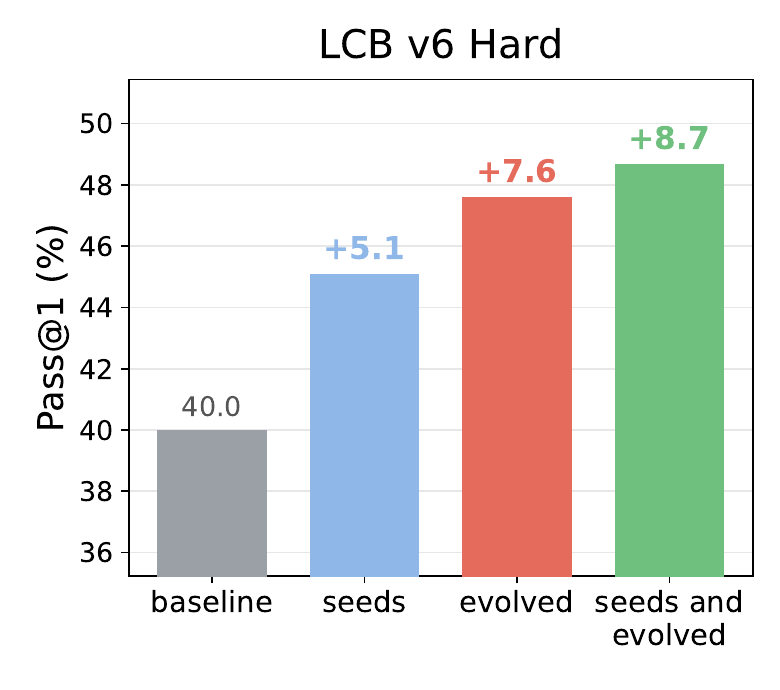}
        \caption{LCB v6 Hard}
        \label{fig:bar_lcb_v6_hard}
      \end{subfigure}
      \hfill
      \begin{subfigure}[t]{0.32\linewidth}
        \centering
        \includegraphics[width=\linewidth]{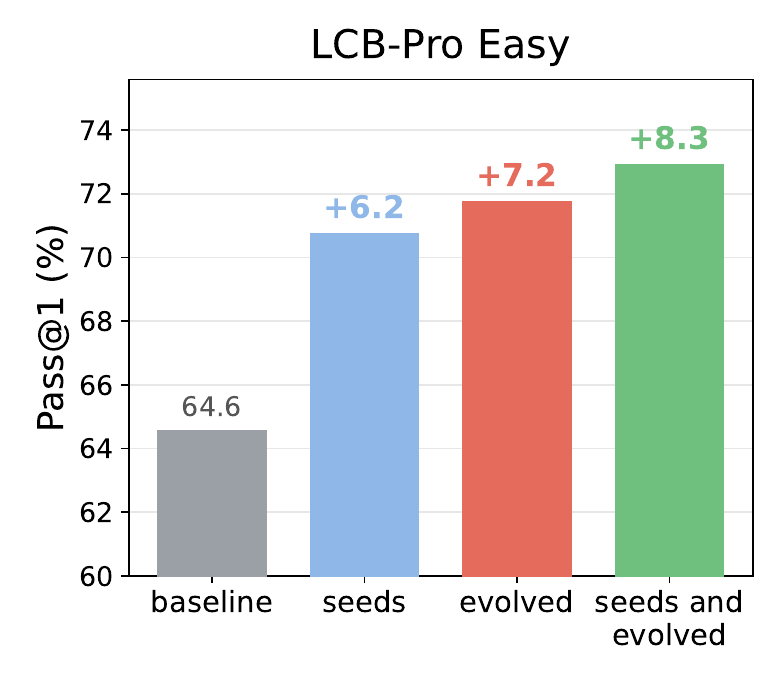}
        \caption{LCB-Pro Easy}
        \label{fig:bar_lcb_pro_easy}
      \end{subfigure}
      \hfill
      \begin{subfigure}[t]{0.32\linewidth}
        \centering
        \includegraphics[width=\linewidth]{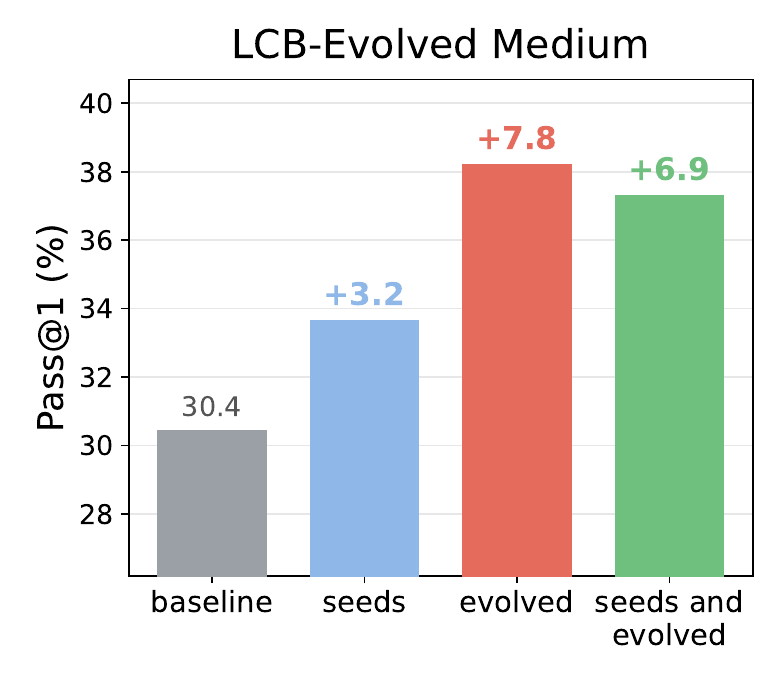}
        \caption{\textsc{LCB-Evolved} Medium}
        \label{fig:bar_lcb_evolved_medium}
      \end{subfigure}
      \caption{Peak observed pass@1 during RL training for each data mixture, compared with the base model. For each evaluation set and data mixture, we report the checkpoint with the highest two-seed mean accuracy observed along the training trajectory; the number above each bar gives the absolute improvement in percentage points over the base model. The truncated \(y\)-axes are used only for readability.}
      \label{fig:best_ckpt_bars}
\end{figure*}

\paragraph{Results on public held-out benchmarks.}
Figures~\ref{fig:acc_curves} and~\ref{fig:best_ckpt_bars} show that evolved tasks provide useful training signal beyond the original seed distribution. Figure~\ref{fig:acc_curves} reports the full training trajectories, while Figure~\ref{fig:best_ckpt_bars} summarizes the peak observed performance of each data mixture during training. On LCB v6 Hard, seed-only RL improves the base model from \(40.0\%\) to \(45.1\%\), whereas evolved-only training reaches \(47.6\%\) and the seed+evolved mixture reaches \(48.7\%\). Thus, incorporating evolved tasks yields an additional \(+2.5\) points for evolved-only training and \(+3.6\) points for the combined mixture over seed-only RL. The same trend holds on LCB-Pro Easy: seed-only training improves from \(64.6\%\) to \(70.8\%\), while evolved-only and seed+evolved training reach \(71.8\%\) and \(72.9\%\), corresponding to additional gains of \(+1.0\) and \(+2.1\) points. In both public held-out settings, the seed+evolved mixture performs best, suggesting that evolved tasks complement the coverage of the original seed distribution while directing learning toward weaknesses not exposed by saturated seeds.

\paragraph{Transfer to an independently evolved benchmark.}
We further evaluate the same RL runs on \textsc{LCB-Evolved} Medium. This benchmark is generated independently by Gemini-3-Flash rather than by \texttt{gpt-oss-20b}, and therefore tests whether self-generated training data transfers beyond the model's own evolved task distribution. We use the Medium split because it remains within the informative difficulty range for \texttt{gpt-oss-20b}: the base model achieves nonzero but far-from-saturated accuracy. Specifically, the base model obtains \(30.45\%\) Pass@1, and seed-only training improves performance modestly to \(33.66\%\). In contrast, training on problems evolved by \texttt{gpt-oss-20b} itself reaches \(38.22\%\), a \(+7.77\)-point gain over the base model and \(+4.56\) points beyond seed-only training; the seed+evolved mixture also improves performance to \(37.32\%\). Unlike the public held-out benchmarks, evolved-only training performs best on \textsc{LCB-Evolved} Medium, indicating that self-generated evolved tasks are especially effective for improving performance on harder evolved-style challenges. Since the evaluation set is produced by a different external evolver, the gains are not simply due to overlap with the training tasks or artifacts of the same evolver.

\paragraph{Closing the self-improvement loop.}
These results support a closed-loop self-improvement interpretation. Starting from seed problems that \texttt{gpt-oss-20b} already solves reliably, \methodname uses inference-time computation to construct verified variants that expose new failures of the current policy. Reinforcement learning then amortizes these self-generated challenges into the model parameters. The evaluation pattern clarifies the role of evolved data: mixing seeds and evolved tasks gives the strongest transfer to future public benchmarks, while evolved-only training gives the largest gain on the independently constructed \textsc{LCB-Evolved} Medium split. Thus, evolved tasks are not only harder evaluation items; they serve as reusable training signal that helps the model improve on difficult coding regimes beyond its original saturated training distribution. We provide the training details in Appendix~\ref{app:training-details}.

\section{Conclusion and Future Work}
\label{sec:conclustion_limitation}

\subsection{Conclusion}
We presented \methodname, a solution-centric framework for turning saturated executable coding tasks into harder, verified challenges. The key idea is to evolve the computation first---by mutating reference solutions---and then recover statements and tests around the evolved solution, keeping task generation grounded in executable semantics. Across LiveCodeBench and SciCode, \methodname produces valid, diverse tasks that substantially reduce target-model pass rates, including for the models that generated them. Building on these results, we curate \textsc{LiveCodeBench-Plus}, a 91-problem benchmark that combines 64 evolved and 27 difficult original LiveCodeBench-v6 tasks and restores meaningful discrimination among frontier coding models. More importantly, evolved tasks are not only useful as harder benchmarks: they also provide actionable training signal. Our RL experiments show that training on evolved problems improves held-out coding performance beyond training on the original seeds alone. This suggests a broader role for benchmark evolution: instead of treating evaluation datasets as static artifacts that inevitably saturate, models can use inference-time computation to discover their own failures, convert those failures into verified training environments, and improve through closed-loop self-play.

\subsection{Future Work}

\paragraph{Scaling closed-loop RL self-improvement.}
Our RL experiments instantiate one round of the self-improvement loop: a model evolves saturated problems into harder verified tasks, trains on them through executable rewards, and improves on held-out coding benchmarks. A natural next step is to scale this into a multi-round process, where the improved model becomes the next evolver and generates a new generation of challenges. Such a loop raises important questions about stability, curriculum design, and diversity. If selection is too narrow, the system may overfit to recurring failure modes; if selection is too aggressive, it may generate tasks that are difficult but uninformative for learning. Developing principled mechanisms for controlling task difficulty, preserving algorithmic diversity, and balancing original, evolved, and newly evolved tasks will be essential for turning self-challenging generation into a scalable training paradigm.

\paragraph{Toward living benchmarks.}
Finally, our results suggest that benchmark construction should move beyond static datasets. Any fixed benchmark will eventually saturate as models improve, especially in executable domains where training signal can be extracted from public tasks. Instead of releasing a benchmark as a one-time artifact, future work could maintain a reproducible evolution pipeline that periodically generates, validates, audits, and calibrates new tasks against current frontier models. Such a living benchmark would make evaluation adaptive to model progress while preserving transparency through versioned releases, held-out tests, and documented validation protocols. More broadly, this would align evaluation and training: the same verified tasks that reveal current model failures can also become the environments used to improve future models.

\newpage
\bibliographystyle{unsrtnat}
\bibliography{main.bib}

\newpage
\appendix

\section{Additional Related Work}
\label{appendix:additional_related_work}

\paragraph{Synthetic coding tasks.}
Synthetic data generation has become a central approach for improving and evaluating LLMs' coding capabilities, especially as human-written programming tasks are expensive to collect and curate at scale. Early work primarily synthesizes instruction-following data: WizardCoder and WizardLM evolve seed instructions into more complex variants~\citep{luo2023wizardcoder,xu2025wizardlm}, while SelfCodeAlign generates instruction data for code alignment~\citep{wei2024selfcodealign}. Other work distills reasoning traces for competitive programming~\citep{ahmad2025opencodereasoning}, synthesizes repository-level bug-fix tasks~\citep{pham2025swe}, or constructs synthetic curricula for RL-based code generation~\citep{sancaktar2026deep}. More recent systems move from instruction synthesis to complete executable tasks. AutoCode uses LLMs as competitive-programming problem setters, generating new statements together with reference solutions and tests~\citep{zhou2025autocode}; X-Coder similarly studies fully synthetic competitive-programming data for training code reasoning models~\citep{wu2026xcoder}. These methods demonstrate that synthetic coding data can improve scale and coverage. However, most of them are still organized as data-generation pipelines for downstream models: they generate instructions, tasks, solutions, or curricula, but do not explicitly require the generated tasks to challenge the generator itself.

\paragraph{Self-play and self-guided improvement.}
A complementary line of work studies model-in-the-loop data generation, where
the model creates new supervision targeted to its evolving capabilities.
Self-Challenging Agents instantiate this idea for tool-use agents, generating
verifiable tasks and training an executor through RL feedback
\citep{zhou2025self}. Self-Guided Self-Play studies a solver--conjecturer loop
for formal theorem proving, introducing a guide role to avoid degenerate or
uninformative conjectures \citep{bailey2026scaling}. More closely related in
motivation, Absolute Zero trains a single model to propose and solve
self-generated code-reasoning tasks, using execution both to construct valid
tasks and to verify solver outputs \citep{zhao2026absolute}. R-Zero instead
co-evolves separate Challenger and Solver models from zero external data: the
Challenger generates mathematical questions near the Solver's capability
boundary, while Solver self-consistency provides pseudo-labels for subsequent
training \citep{huang2026rzeroselfevolvingreasoningllm}. In code generation,
related work studies iterative code-centric data synthesis through
Coder--Reviewer feedback \citep{sun2025codeevo}, self-improvement from
model-generated solutions \citep{zhang2026embarrassingly}, and adversarial
co-evolution between code-generation and test-generation models
\citep{wang2026codea1}. Together, these works show that models can generate
renewable training signal by targeting their own current weaknesses.

Our work shares this self-challenging motivation, but targets a different
object: complete benchmark items evolved from existing coding problems.
Rather than generating free-form training questions from scratch, \methodname
mutates an executable reference solution of a real seed task and derives a new
statement and test suite around the evolved computation. Each accepted task is
therefore required to be well specified, executable under the original
benchmark harness, and empirically harder for a panel of target models,
including the evolver itself. This design makes the output suitable not only
as self-generated RL data, but also as an evolved benchmark for evaluating
frontier coding models.

\paragraph{Self-evolving algorithms for LLM agents.}
Evolution-based algorithms have emerged as a general in-context strategy for solving difficult but verifiable optimization problems. They typically combine LLM-generated mutations with automatic evaluators, using feedback from candidate performance to guide subsequent generations. This paradigm has been applied to prompt and context optimization~\citep{fernando2023promptbreeder,guo2023evoprompt,agrawal2025gepa,zhang2025agentic}, program and algorithm discovery~\citep{romera2024funsearch,novikov2025alphaevolve,shojaee2025llm,cheng2025barbarians}, and adaptive evolutionary search, including diversity-driven program evolution, meta-evolution, and long-horizon progress-aware optimization~\citep{lange2025shinkaevolve,jiang2026deltaevolve,yan2026pacevolve,cemri2026adaevolve,liu2026evox}. Related work also studies automated or self-evolving agent systems~\citep{hu2024automated,wu2025evolveR} and open-source evolutionary coding agents~\citep{cao2026codeevolve}.

Our framework adopts the same closed-loop search principle, but applies it to benchmark generation rather than solution optimization. Instead of evolving better prompts, programs, or agents for a fixed objective, \methodname evolves the objective itself: harder executable coding tasks with reference solutions and tests. This shift is important for self-improvement. In conventional evolutionary optimization, the evaluator is fixed and the goal is to find a better solution; in \methodname, the evaluator is used to discover new tasks that expose the model's current weaknesses, turning inference-time search~\citep{wu2024inference} into reusable benchmark and training data.

\section{Pseudocode for \methodname}
\label{appendix:pseudocode}

\paragraph{Notation for Algorithm~\ref{alg:pseudocode}.}
\label{app:alg-notation}
Each task is $I=(S,C,T,E)$, consisting of a statement, reference code, hidden
tests, and execution harness. The panel $\Pi=\{\pi_j\}_{j=1}^{J}$ contains the
target solvers, and $\hat C_{j,k}\sim\pi_j(S)$ is the $k$-th solution attempt
from solver $\pi_j$. The map $\phi$ converts lower empirical accuracy to higher
difficulty level. The operators $q_\theta$, $w_\theta$, and $g_\theta$ are the
mutator, statement writer, and test generator. For seed $I_i$, $L_i$ is the
accepted lineage, $m_i$ stores local history such as accepted/rejected
mutations, repairs, scores, and target error patterns, and $G$ stores accepted
mutation ideas across seeds for diversity guidance. The context $h_{i,b}$
packages these memories and difficulty levels for proposal step $b$; $r(I')$
is the validation predicate; and $A(I')$ is the final acceptance predicate.

\begin{algorithm}
\caption{\methodname: Solution-Centric Evolution}
\label{alg:pseudocode}
\begin{algorithmic}[1]
\Require Seed benchmark $\mathcal{D}=\{I_i=(S_i,C_i,T_i,E_i)\}_{i=1}^n$,
target solver panel $\Pi=\{\pi_j\}_{j=1}^J$, attempts per solver $K$, evolution
budget $B$, minimum level gain $\Delta$
\Ensure Evolved benchmark $\mathcal{D}'$
\Statex $V_E(\hat C,T)=1$ iff $\hat C$ passes all tests $T$ under harness $E$.
\Statex $a(I;\Pi,K)=\frac{1}{JK}\sum_{j=1}^{J}\sum_{k=1}^{K}V_E(\hat C_{j,k},T)$ and $\ell(I)=\phi(a(I;\Pi,K))$.
\Statex $L_i,m_i,G,h_{i,b},r(I'),A(I')$ are defined in Appendix~\ref{app:alg-notation}.
\State $\mathcal{D}' \gets \emptyset$, $G \gets \emptyset$
\ForAll{$I_i=(S_i,C_i,T_i,E_i)\in\mathcal{D}$}
    \State $a_i \gets a(I_i;\Pi,K)$, $\ell_i \gets \phi(a_i)$, $L_i \gets [\,]$, $m_i \gets \emptyset$
    \For{$b=1,\ldots,B$}
        \Statex \textsc{Parent/context:} latest accepted child, not sampled parent
        \State $I_p=(S_p,C_p,T_p,E_p)\gets \mathrm{Last}(L_i)$ if $L_i\ne\emptyset$, else $I_i$;
        $h_{i,b}\gets(m_i,G,\ell_i,\ell(I_p))$
        \Statex \textsc{Proposer:} evolve solution, then derive statement/tests
        \State $C'\sim q_\theta(\cdot\mid C_p,h_{i,b})$; materialize examples by executing $C'$
        \State $S'\sim w_\theta(\cdot\mid C')$, $T'\sim g_\theta(\cdot\mid S',C')$; set $I'=(S',C',T',E_i)$
        \Statex \textsc{Validation:} executable consistency with bounded repair
        \State $r(I')\gets\mathbf{1}\{\mathrm{Consistent}(S',C',T',E_i)\}$; repair and recompute $r(I')$ if needed
        \If{$r(I')=0$} \State record rejection in $m_i$ and continue \EndIf
        \Statex \textsc{Difficulty/selection:} target-panel failure after validation
        \State $a'\gets\frac{1}{JK}\sum_{j=1}^{J}\sum_{k=1}^{K}V_{E_i}(\hat C'_{j,k},T')$,
        $\ell'\gets\phi(a')$
        \State $A(I')\gets r(I')\wedge[\ell'\ge\ell(I_p)]\wedge[\ell'\ge\ell_i+\Delta]
        \wedge\neg\mathrm{Artificial}(I')$
        \If{$A(I')=1$}
            \State $\mathcal{D}'\gets\mathcal{D}'\cup\{I'\}$; append $I'$ to $L_i$; update $m_i$ and $G$
        \Else
            \State Record rejection reason and observed failures in $m_i$
        \EndIf
    \EndFor
\EndFor
\State \Return $\mathcal{D}'$
\end{algorithmic}
\end{algorithm}

\section{Validation and Repair Details}
\label{app:validation}

\subsection{LiveCodeBench brute-force triangulation}
\label{app:lcb_validation}

For LiveCodeBench, \methodname applies a benchmark-specific self-validation stack before target-model evaluation. It synthesizes an independent brute-force solver from the statement alone and checks public examples using a three-way vote among the reference solution, the brute-force solver, and a natural-language oracle that sees only the problem statement. Agreement passes the candidate; disagreements trigger targeted repair of the brute-force solver, reference solution, or task specification. The vetted brute-force solver is then run on generated hidden tests where feasible, using concrete output disagreements as counterexamples for repair while treating large-case timeouts as brute-force infeasibility. Candidates are rejected if the validation stack cannot resolve the inconsistency within a shared repair budget of three attempts, with each repair action, regardless of type, consuming one attempt.

\subsection{SciCode statement-faithfulness validation}
\label{app:scicode_validation}

SciCode tasks do not naturally support brute-force validation because they are scientific function-level problems with assertion-based tests and domain-specific conventions. We therefore use a best-of-\(N\) statement-faithfulness check. After generating the statement, reference solution, and hidden assertion tests, the evaluator model solves the task from the statement alone. Each alternate solution is executed against the generated tests, and we keep the best pass rate over \(N\) attempts. We set a threshold of 0.5, and candidates below this threshold are treated as underspecified; the pipeline revises the statement and reruns the check against the same tests. If the revised task still fails within the shared repair budget, the candidate is rejected. This gate is intended to detect specification gaps rather than certify scientific correctness. Same-model self-play may share numerical blind spots with the reference solution, but if a capable solver cannot reproduce the expected behavior from the statement after several attempts, the task likely omits an important convention, assumption, or return-contract detail.

\section{Training Details}
\label{app:training-details}

We fine-tune the open-weight \texttt{openai/gpt-oss-20B} model with on-policy
reinforcement learning on LiveCodeBench-style coding problems. Training is
performed through the Tinker RL service \citep{tml2025tinker} with LoRA adapters on
the policy; the reference policy and the sampler share the same base weights.
All rollouts execute generated programs in an isolated cloud sandbox so that
the reward signal is grounded in real test-case outcomes rather than a learned
reward model.

\subsection{Optimization and Model Configuration}
\label{app:hparams}

The policy uses LoRA adapters of rank $r{=}32$ over the attention and MLP
projections of the base model. We optimize with a constant learning rate of
$\eta{=}1{\times}10^{-5}$ and a single optimizer substep per batch
($K{=}1$). We do not apply a KL penalty to the reference policy
($\beta_{\text{KL}}{=}0$) and instead rely on the LoRA bottleneck and the
group-relative advantages (Sec.~\ref{app:rl-objective}) to keep the policy
close to the base model. Conversations are formatted with the
\texttt{gpt\_oss\_medium\_reasoning} chat renderer, which preserves the
model's native thinking--answer structure.

\begin{table*}[ht]
\centering
\small
\begin{tabular}{lll}
\toprule
\textbf{Component} & \textbf{Hyperparameter} & \textbf{Value} \\
\midrule
Base model        & \texttt{model\_name}              & \texttt{openai/gpt-oss-20B} \\
LoRA              & \texttt{lora\_rank}               & $32$ \\
Renderer          & \texttt{renderer\_name}           & \texttt{gpt\_oss\_medium\_reasoning} \\
Optimizer         & \texttt{learning\_rate}           & $1{\times}10^{-5}$ \\
                  & \texttt{kl\_penalty\_coef}        & $0.0$ \\
                  & \texttt{num\_substeps}            & $1$ \\
Rollouts          & \texttt{group\_size} ($G$)        & $16$ \\
                  & \texttt{groups\_per\_batch} ($B$) & $64$ \\
                  & \texttt{seed}                     & $42$ and $43$ \\
Context (train)   & \texttt{max\_tokens}              & $24{,}000$ \\
                  & \texttt{max\_trajectory\_tokens}  & $26{,}500$ \\
Context (eval)    & \texttt{test\_max\_tokens}        & $30{,}000$ \\
                  & \texttt{test\_max\_trajectory\_tokens} & $32{,}500$ \\
Environment       & \texttt{sandbox\_backend}         & \texttt{modal} \\
                  & per-test \texttt{timeout}         & $6$\,s \\
Reward shaping    & \texttt{format\_coef} ($\lambda$) & $0.1$ \\
                  & \texttt{context\_overflow\_reward}& $-0.1$ \\
\bottomrule
\end{tabular}
\caption{Training configuration used for all RL runs. Train/eval context caps
differ because the policy needs more room at evaluation time to produce a
valid final answer when reasoning chains are longest.}
\label{tab:hparams}
\end{table*}

\subsection{RL Objective and Batch Construction}
\label{app:rl-objective}

For each task $\tau$ in a training batch, we sample $G{=}16$ independent
trajectories $\{x^{(g)}\}_{g=1}^{G}$ from the current policy $\pi_\theta$ at
nucleus sampling temperature defaults. A batch consists of $B{=}64$ such
groups, yielding $B \cdot G {=} 1024$ trajectories per gradient step. We use
group-relative advantages: for each task, the scalar reward of every
trajectory is centered against the mean reward of its own group,

\begin{equation}
A^{(g)}_\tau \;=\; r(x^{(g)}_\tau) \;-\; \frac{1}{G}\sum_{g'=1}^{G} r(x^{(g')}_\tau),
\label{eq:advantage}
\end{equation}

so that easy tasks (where most rollouts pass) and hard tasks (where most
rollouts fail) contribute meaningful gradient signal without an additional
value baseline. Trajectories are then trained against the policy-gradient
loss with clipped importance weights, in the standard on-policy form used by
Tinker's RL trainer.

\subsection{Reward Design}
\label{app:reward}

Rewards are computed once per trajectory, after the entire conversation has
finished. The grader extracts the \emph{last} fenced code block from the final
assistant turn and submits it to the cloud sandbox along with the task's
LiveCodeBench-format tests.

Let $c \in \{0,1\}$ indicate whether the extracted program passes
\emph{every} test case under a per-test wall-clock limit of $T{=}6$\,s, and
let $f \in \{0,1\}$ indicate whether the response contains at least one
fenced code block. The episode reward is

\begin{equation}
r \;=\; c \;+\; \lambda\,(f - 1),
\qquad \lambda = 0.1.
\label{eq:reward}
\end{equation}

Equation~\ref{eq:reward} has three intended properties. (i) A fully correct
solution receives $r{=}1$. (ii) An incorrect but well-formatted attempt
receives $r{=}0$, so the policy is not punished for trying. (iii) A response
without an extractable code block receives $r{=}-0.1$, providing a small
shaping signal that prevents the policy from collapsing into pure chain-of-thought without ever emitting code. We deliberately keep $\lambda$ small so
that format shaping never dominates the correctness signal.

\paragraph{Sandbox grading.} Stdin/stdout problems are run as a real Python
subprocess inside the sandbox, matching the LiveCodeBench harness; functional
problems use the in-process call-based path. Output comparison applies, in
order, exact-match (after stripping trailing whitespace), numeric match with
relative tolerance $10^{-6}$, and a set-of-tokens fallback for unordered
outputs. Per-test failures (Wrong Answer, Time Limit Exceeded, Runtime
Error) all count as $c{=}0$, regardless of which test failed first.

\paragraph{Trajectory-level penalties.} Trajectories whose token budget is
exhausted before a final answer is produced receive
$r_{\text{overflow}}{=}-0.1$, identical in magnitude to the format penalty.
This prevents the policy from learning to stall indefinitely in the reasoning
channel.

\subsection{Training Dynamics}
\label{subsec:training_dynamics}

\begin{figure}[ht]
\centering
\includegraphics[width=0.95\linewidth]{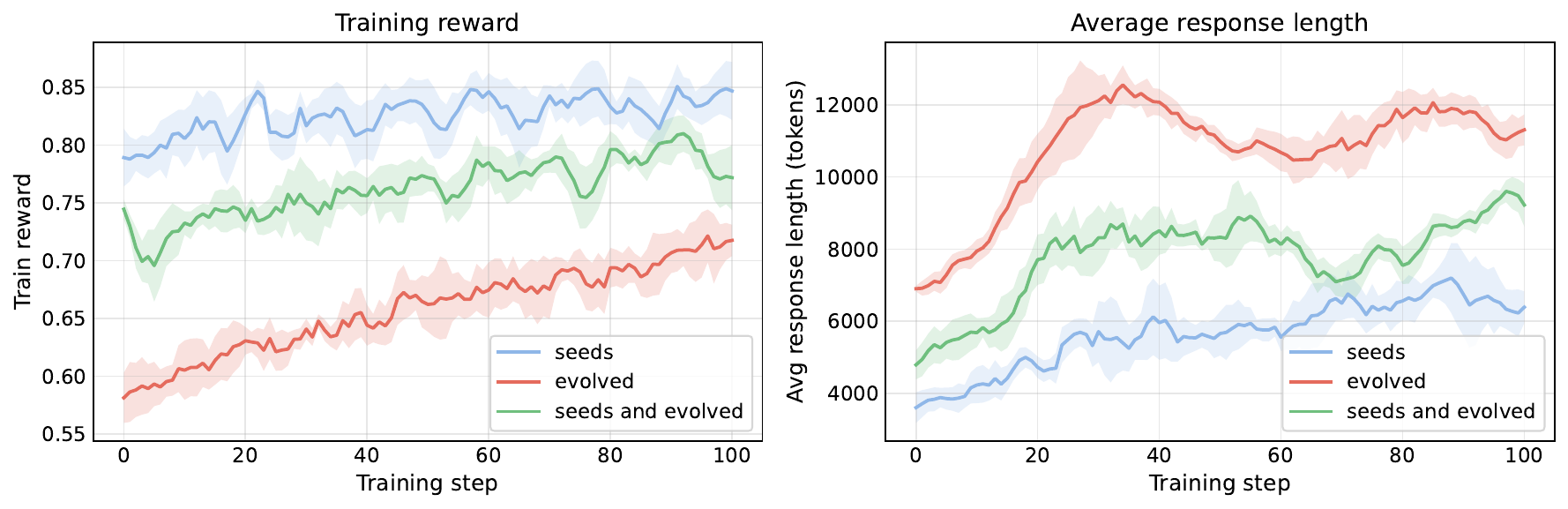}
\caption{Training reward (left) and average response length in tokens
(right), for the three data conditions
(seeds, evolved, seeds and evolved). Solid lines are the mean of two random
seeds per condition; shaded bands are \(\pm 1\) standard deviation
across the two seeds. A boxcar smoother of width \(5\) is applied to both the
mean and standard-deviation curves. The reward panel reflects
Eq.~\ref{eq:reward}; the maximum attainable per-trajectory reward is \(1.0\)
and the minimum is \(-0.1\).}
\label{fig:train-curves}
\end{figure}

Figure~\ref{fig:train-curves} reports the on-policy training reward and average response length for the three RL data mixtures. The reward curves start at different levels because the datasets have different difficulty: seed problems are easier for the base policy and therefore begin with higher reward, while evolved problems are intentionally harder and start substantially lower. This gap is expected and reflects the construction of \methodname: evolved tasks are selected to expose failures of the current model rather than to maximize initial reward. Across training, all mixtures improve, but with different dynamics. Seed-only training begins high and quickly saturates, suggesting limited remaining learning signal. Evolved-only training starts from the lowest reward but rises steadily throughout training, indicating that the evolved distribution provides nontrivial gradient signal over many updates. The combined seed+evolved mixture lies between the two at initialization and also improves consistently, balancing easier problems that stabilize training with harder evolved tasks that continue to drive learning.

\subsection{Compute and Reproducibility}
\label{subsec:compute_reproducibility}

Each RL run takes approximately \(40\) hours and costs about \(\$800\) in Tinker credits. All runs use the Tinker service for policy execution and gradient updates, while the client-side training loop runs on a single CPU host. Code execution is delegated to Modal-hosted sandboxes, so throughput is primarily determined by generated-program execution and test-case latency rather than local compute. We use two independent random seeds for each data mixture and keep the hyperparameters fixed across all runs.

\section{Evolution Configurations and Hyperparameters}
\label{app:evolution_config}
The default configurations for our LiveCodeBench and SciCode experiments are summarized in Table~\ref{tab:configs}. The table reports the model settings, acceptance criteria, repair and validation budgets, test-generation parameters, and memory mechanisms used by our evolution pipeline.

\newcolumntype{L}[1]{>{\raggedright\arraybackslash}p{#1}}
\newcommand{\cfg}[1]{\begingroup\urlstyle{tt}\scriptsize\nolinkurl{#1}\endgroup}
\newcommand{\dash}{--}

\begin{table*}[t]
\centering
\scriptsize
\setlength{\tabcolsep}{3pt}
\renewcommand{\arraystretch}{1.12}
\begin{tabularx}{\linewidth}{@{}
  L{0.25\linewidth}
  L{0.40\linewidth}
  L{0.16\linewidth}
  L{0.16\linewidth}
@{}}
\toprule
\textbf{Config name} & \textbf{Purpose} & \textbf{LCB} & \textbf{SciCode} \\
\midrule

\multicolumn{4}{@{}l}{\textbf{Model and evaluation}} \\
\midrule
\cfg{target_eval_k} & Attempts per target model for evolved problems. &
\cfg{4} & \cfg{4} \\
\cfg{temperature} & Sampling temperature for generation. &
\cfg{0.8} & \cfg{0.8} \\
\cfg{timeout} & LLM request timeout in seconds. &
\cfg{600} & \cfg{600} \\

\midrule
\multicolumn{4}{@{}l}{\textbf{Difficulty and acceptance}} \\
\midrule
\cfg{allowed_seed_levels} & Seed difficulty levels allowed as starting points. &
\cfg{[1, 2]} & \cfg{[1, 2]} \\
\cfg{accept_min_level_gain} & Minimum difficulty-level gain over the seed. &
\cfg{1} & \cfg{1} \\
\cfg{accept_target_level} & Stop once this accepted difficulty level is reached. &
\cfg{5} & \cfg{5} \\
\cfg{max_iters_per_seed} & Maximum evolution iterations per seed. &
\cfg{10} & \cfg{10} \\
\cfg{max_accepted_per_seed} & Maximum accepted evolved items per seed. &
\cfg{3} & \cfg{3} \\
\cfg{no_improve_patience} & Stop evolution after consecutive non-accepted iterations. &
\cfg{4} & \cfg{4} \\

\midrule
\multicolumn{4}{@{}l}{\textbf{Mutation and repair}} \\
\midrule
\cfg{max_candidate_repairs} & Shared repair budget per candidate. &
\cfg{3} & \cfg{3} \\
\cfg{spec_retry_attempts} & Statement/spec revision attempts. &
\cfg{2} & \cfg{2} \\
\cfg{enable_test_repair} & Allow regenerating weak or malformed tests. &
\cfg{true} & \cfg{true} \\
\cfg{test_repair_attempts} & Test repair attempts. &
\cfg{2} & \cfg{2} \\

\midrule
\multicolumn{4}{@{}l}{\textbf{SciCode-specific faithfulness checks}} \\
\midrule
\cfg{faithfulness_check_N} & Alt-solver attempts for statement-faithfulness check. &
\dash & \cfg{3} \\
\cfg{faithfulness_min_pass_rate} & Best-of-$N$ pass-rate floor for a faithful statement. &
\dash & \cfg{0.5} \\
\cfg{faithfulness_max_repair} & Spec-revision retries on faithfulness failure. &
\dash & \cfg{2} \\

\midrule
\multicolumn{4}{@{}l}{\textbf{Test generation and execution}} \\
\midrule
\cfg{program_gen_small_inputs} & Number of small generated inputs. &
\cfg{5} & \cfg{2} \\
\cfg{program_gen_medium_inputs} & Number of medium generated inputs. &
\cfg{5} & \cfg{2} \\
\cfg{program_gen_large_inputs} & Number of large generated inputs. &
\cfg{5} & \cfg{2} \\
\cfg{program_gen_stress_inputs} & Number of stress generated inputs. &
\cfg{3} & \dash \\
\cfg{execution_timeout} & Per-test reference/target execution timeout. &
\cfg{6.0} & \cfg{30.0} \\
\cfg{max_output_bytes} & Maximum captured stdout bytes. &
\cfg{1000000} & \cfg{1000000} \\

\midrule
\multicolumn{4}{@{}l}{\textbf{Memory and diversity}} \\
\midrule
\cfg{memory_raw_window} & Raw recent iteration records retained in local memory. &
\cfg{10} & \cfg{5} \\
\cfg{memory_digest_recent_k} & Recent raw records shown alongside digest. &
\cfg{3} & \cfg{3} \\
\cfg{judge_near_duplicate_check} & Judge rejects near-duplicates within a seed lineage. &
\cfg{true} & \cfg{true} \\
\cfg{global_memory_max_entries} & Max global accepted entries shown to mutator. &
\cfg{50} & \cfg{20} \\
\cfg{global_memory_max_chars} & Character cap for global memory block. &
\cfg{6000} & \cfg{6000} \\

\bottomrule
\end{tabularx}
\caption{Default evolution configurations for LCB and SciCode.}
\label{tab:configs}
\end{table*}

\section{Human Evaluation: Full Breakdown}
\label{app:human_eval_full}

\begin{figure}[t]
\centering
\begin{subfigure}[t]{0.48\linewidth}
  \centering
  \includegraphics[width=\linewidth]{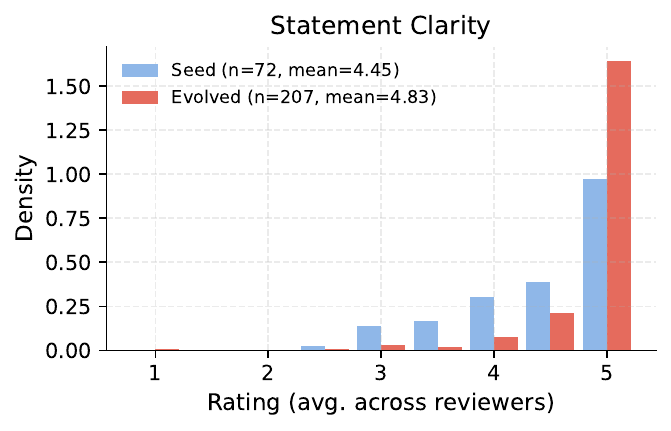}
  \caption{Clarity.}
  \label{fig:appx_he_clarity}
\end{subfigure}\hfill
\begin{subfigure}[t]{0.48\linewidth}
  \centering
  \includegraphics[width=\linewidth]{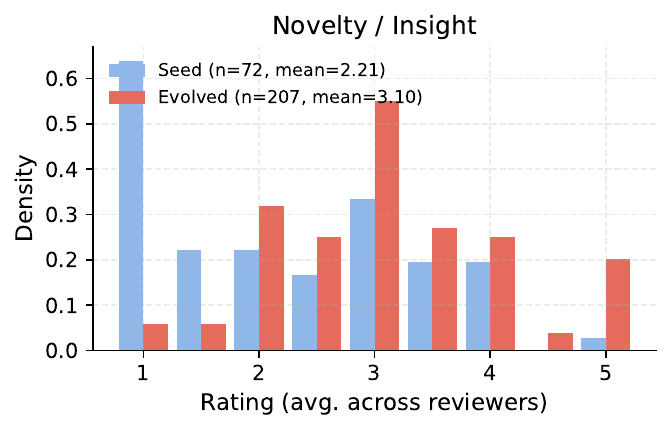}
  \caption{Novelty / insight.}
  \label{fig:appx_he_novelty}
\end{subfigure}

\vspace{4pt}

\begin{subfigure}[t]{0.48\linewidth}
  \centering
  \includegraphics[width=\linewidth]{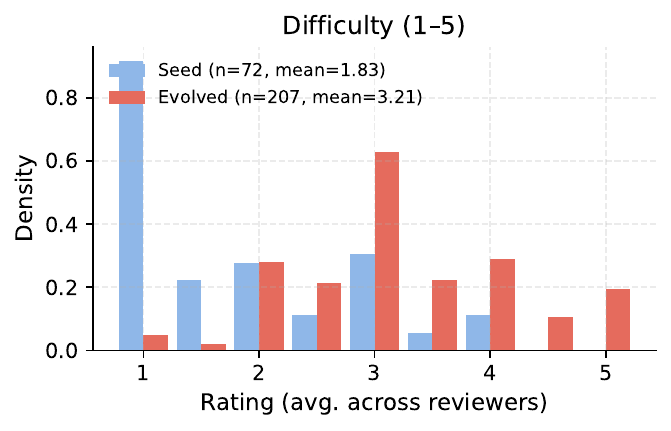}
  \caption{Difficulty.}
  \label{fig:appx_he_difficulty}
\end{subfigure}\hfill
\begin{subfigure}[t]{0.48\linewidth}
  \centering
  \includegraphics[width=\linewidth]{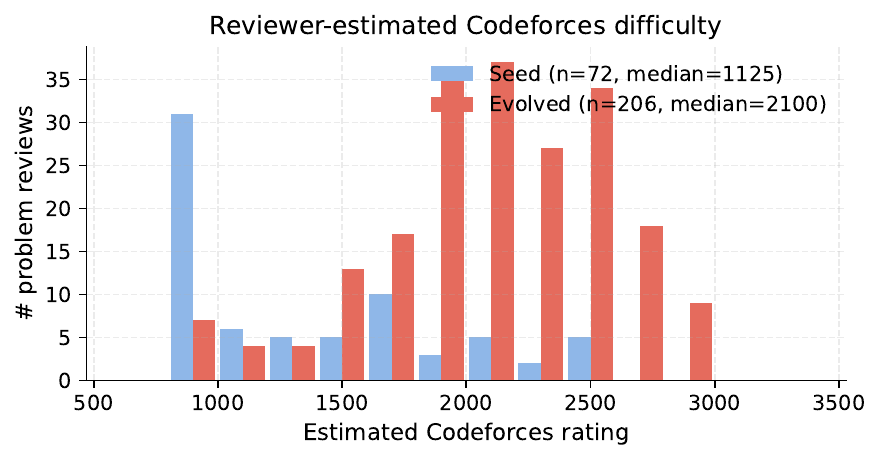}
  \caption{Estimated Codeforces rating.}
  \label{fig:appx_he_cf}
\end{subfigure}

\vspace{4pt}

\begin{subfigure}[t]{0.48\linewidth}
  \centering
  \includegraphics[width=\linewidth]{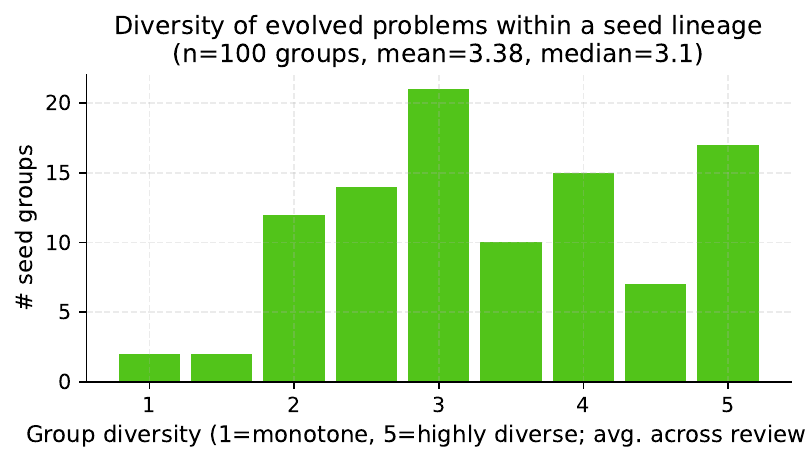}
  \caption{Group-level diversity.}
  \label{fig:appx_he_div}
\end{subfigure}\hfill
\begin{subfigure}[t]{0.48\linewidth}
  \centering
  \includegraphics[width=\linewidth]{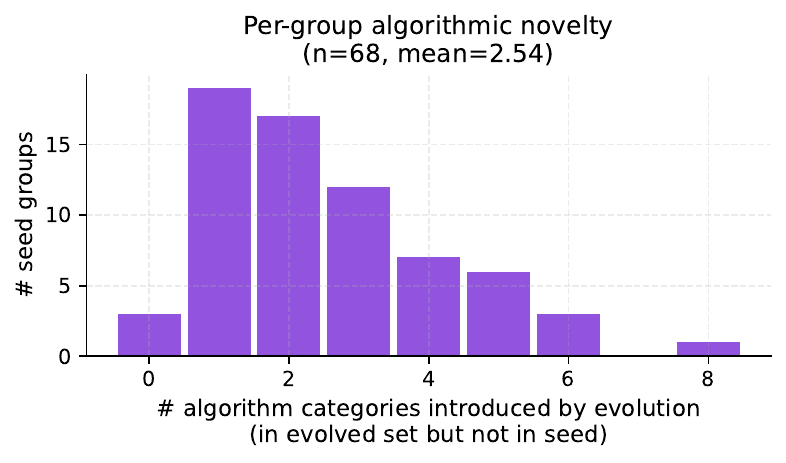}
  \caption{New categories per group.}
  \label{fig:appx_he_newcats}
\end{subfigure}

\caption{
Human-evaluation distributions. The evolved problems are rated as more novel
and more difficult than their seeds, with median estimated Codeforces rating
increasing from $1125$ to $2100$. At the group level, evolved lineages have
mean diversity $3.38/5$ and introduce an average of $2.54$ new algorithm
categories per group, with at least one new category in $95.6\%$ of groups.
}
\label{fig:appx_human_eval_grid}
\end{figure}

\begin{figure}[t]
\centering
\includegraphics[width=0.85\linewidth]{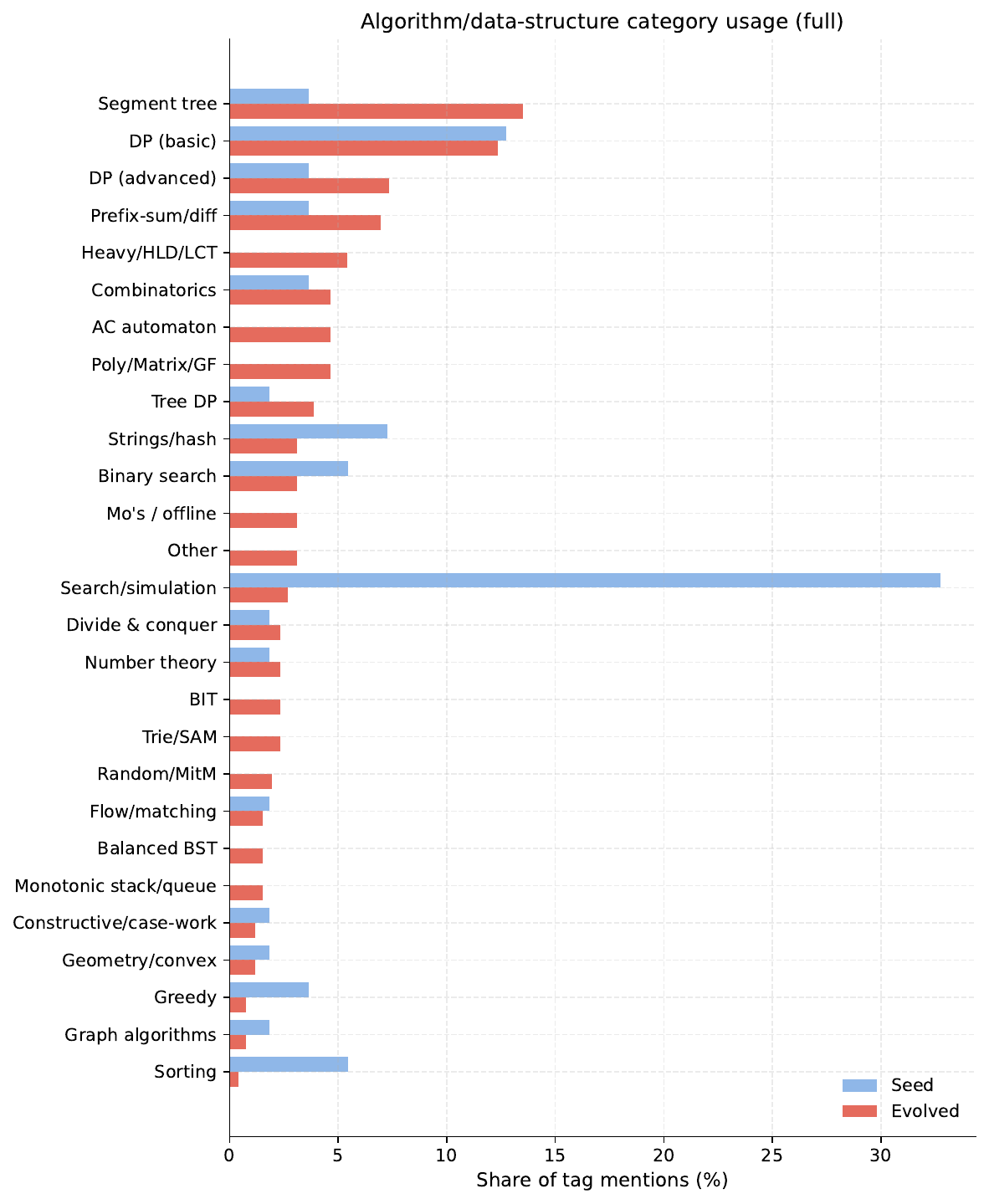}
\caption{
Full algorithm-category distribution for seed and evolved problems, including
all categories with at least three combined mentions. The main-body
Figure~\ref{fig:human_eval_tags} shows the eleven categories with the largest
absolute seed-to-evolved share shift.
}
\label{fig:appx_human_eval_full_tags}
\end{figure}

This appendix provides the full human-evaluation breakdown for
Section~\ref{paragraph:human_eval} and Figure~\ref{fig:human_eval_tags},
including distributional statistics (Figure~\ref{fig:appx_human_eval_grid})
and the complete algorithm-category distribution
(Figure~\ref{fig:appx_human_eval_full_tags}).

\paragraph{Review protocol.}
We use a blinded expert-review protocol to reduce positional, model, and
post-hoc filtering biases. Six competitive-programming experts
(Codeforces grandmaster / IOI / ICPC level) review groups of 2 to 4 anonymized
problems with opaque identifiers and are not told which problem is the seed,
which are evolved variants, or which model generated them. Reviewers certify
that they do not use generative AI during evaluation. For each problem, they
rate clarity, novelty, and difficulty on a \(1\)--\(5\) scale, estimate a
Codeforces rating, list required algorithms and
data structures, and provide short written justifications. Groups are assigned
to two reviewers when possible: \(65\%\) of problems receive two
independent reviews. We aggregate numeric ratings by averaging per problem, and aggregate algorithm tags by taking the union after normalizing synonyms and variants into a controlled vocabulary. This protocol makes the reported statistics
primarily reflect problem-level variation rather than a single reviewer or
post-hoc selection.

\paragraph{Results.}
Figure~\ref{fig:appx_human_eval_grid} shows a consistent shift from the seed
problems to the evolved problems. The evolved problems are rated substantially
harder (panel~c; \(1.83\to3.21\)), and the estimated Codeforces ratings
(panel~d) move from a concentrated pupil/specialist range around \(1100\) to a
broader distribution centered near \(2100\). They are also judged more novel
(panel~b; \(2.21\to3.10\)), with far fewer cases rated as close variants of
the seed. Importantly, this increase in difficulty does not come from making
the statements less clear: clarity improves from \(4.45\) to \(4.83\), with most
evolved problems rated \(4\) or \(5\). Finally, panels~e--f and
Figure~\ref{fig:appx_human_eval_full_tags} show that the evolved problems
broaden the algorithmic coverage: \(95.6\%\) of lineages introduce at least one
new algorithmic category, with \(2.54\) new categories per group on average,
and the total number of observed categories increases from \(19\) to \(30\).
Overall, the human study supports the main claim that \methodname produces
problems that are harder and diverse while remaining well specified.

\FloatBarrier

\section{Examples of Evolved Benchmark Items}
In this section, we provide examples of successfully evolved frontier tasks for LiveCodeBench and SciCode, and compare them with their seed problems.

\subsection{LiveCodeBench Examples}

\definecolor{EvoBlue}{HTML}{233654}
\definecolor{EvoGreen}{HTML}{0B7A55}
\definecolor{EvoPurple}{HTML}{6A4C93}
\definecolor{EvoOrange}{HTML}{A85D22}
\definecolor{EvoLightGray}{HTML}{F6F8FB}
\definecolor{EvoBorderGray}{HTML}{D7DDE7}

\captionsetup[figure]{font=small,skip=4pt}

\newcommand{\evopill}[2]{%
  \tcbox[
    on line,
    colback=#1!78!black,
    colframe=#1!78!black,
    boxrule=0pt,
    arc=8pt,
    left=6pt,right=6pt,top=1pt,bottom=1pt
  ]{\color{white}\scriptsize\bfseries #2}%
}

\newcommand{\evocardheader}[3]{%
\begin{tcolorbox}[
  enhanced,
  colback=#1,
  colframe=#1,
  boxrule=0pt,
  arc=5pt,
  left=8pt,right=8pt,top=5pt,bottom=5pt
]
\begin{tabularx}{\linewidth}{@{}>{\RaggedRight\arraybackslash}Xr@{}}
{\color{white}\bfseries\large #2} & \evopill{#1}{#3}
\end{tabularx}
\end{tcolorbox}
\vspace{0.12em}
}

\newcommand{\evosechead}[2]{%
  \vspace{0.10em}
  \noindent{\color{#1}\rule{2pt}{0.85ex}}\hspace{0.4em}%
  {\bfseries\scriptsize\MakeUppercase{#2}}\par
  \vspace{0.03em}
}

\newtcolorbox{evoproblemcard}{%
  enhanced,
  colback=white,
  colframe=EvoBorderGray,
  boxrule=0.65pt,
  arc=7pt,
  left=8pt,right=8pt,top=7pt,bottom=7pt,
  before skip=0.30em,
  after skip=0.30em
}

\newtcolorbox{evosoftbox}{%
  enhanced,
  colback=EvoLightGray,
  colframe=EvoBorderGray,
  boxrule=0.45pt,
  arc=5pt,
  left=5pt,right=5pt,top=4pt,bottom=4pt,
  before upper={\RaggedRight\setlength{\emergencystretch}{2em}}
}

\newcommand{\evoex}[2]{%
  \noindent
  \begin{tabularx}{\linewidth}{@{}>{\RaggedRight\arraybackslash\ttfamily\tiny}X
                                  >{\RaggedLeft\arraybackslash}p{0.15\linewidth}@{}}
  #1 & $\rightarrow$ \texttt{#2}
  \end{tabularx}\par\vspace{0.12em}
}
\newcommand{\evoexample}[2]{%
  \noindent
  \begin{minipage}[t]{0.78\linewidth}
    \ttfamily\tiny #1
  \end{minipage}\hfill
  \begin{minipage}[t]{0.18\linewidth}
    \raggedleft $\rightarrow$ \texttt{#2}
  \end{minipage}\par\vspace{0.3em}
}

\newtcolorbox{evosuitebox}{%
  enhanced,
  breakable,
  colback=white,
  colframe=EvoBorderGray,
  boxrule=0.8pt,
  arc=7pt,
  left=8pt,right=8pt,top=8pt,bottom=8pt,
  pad at break=1mm
}

\newcommand{\evopagebreak}{\par\vspace{0.2em}\pagebreak[4]\vspace{0.2em}}

\begin{evoproblemcard}
\RaggedRight
\evocardheader{EvoBlue}{LCB Seed: Copy Arrays}{decision/counting $\cdot$ O(N)}

\evosechead{EvoBlue}{Task}
Given \texttt{original} and interval bounds $\texttt{bounds}[i]=[u_i,v_i]$,
count arrays \texttt{copy} such that
\[
\texttt{copy}[i]-\texttt{copy}[i-1]
=
\texttt{original}[i]-\texttt{original}[i-1],
\qquad 1 \le i < n,
\]
with $u_i \le \texttt{copy}[i] \le v_i$ for all $i$.
The key observation is
\[
\texttt{copy}[i]=\texttt{original}[i]+d,
\]
so feasibility reduces to intersecting valid intervals for one global offset $d$.

\vspace{0.1em}
\begin{minipage}[t]{0.34\linewidth}
\evosechead{EvoBlue}{Constraints}
\begin{evosoftbox}
\scriptsize
\[
2 \le n \le 10^5
\]
\[
1 \le \texttt{original}[i] \le 10^9
\]
\[
1 \le u_i \le v_i \le 10^9
\]
\end{evosoftbox}
\end{minipage}\hfill
\begin{minipage}[t]{0.62\linewidth}
\evosechead{EvoBlue}{Examples}
\begin{evosoftbox}
\scriptsize
\evoex{original=[1,2,3,4]; bounds=[[1,2],[2,3],[3,4],[4,5]]}{2}
\evoex{original=[1,2,3,4]; bounds=[[1,10],[2,9],[3,8],[4,7]]}{4}
\evoex{original=[1,2,1,2]; bounds=[[1,1],[2,3],[3,3],[2,3]]}{0}
\end{evosoftbox}
\end{minipage}
\end{evoproblemcard}

\vspace{0.35em}

\begin{evoproblemcard}
\RaggedRight
\evocardheader{EvoGreen}{Evolved Problem: XOR-Linked Sequence}{digit DP $\cdot$ O(N bits)}

\evosechead{EvoGreen}{Task}
The evolved task replaces additive differences with XOR differences:
\[
\texttt{copy}[i]\oplus \texttt{copy}[i-1]
=
\texttt{original}[i]\oplus \texttt{original}[i-1],
\qquad 1 \le i < n.
\]
Count non-negative $x=\texttt{copy}[0]$ such that all entries satisfy
their bounds. Since
\[
\texttt{copy}[i]=x\oplus p_i,
\qquad
p_i=\texttt{original}[i]\oplus \texttt{original}[0],
\]
the task becomes counting $x$ such that
\[
u_i \le x\oplus p_i \le v_i
\qquad \text{for all } i.
\]
Unlike arithmetic intervals, these XOR-induced feasible sets are not contiguous.

\vspace{0.1em}
\begin{minipage}[t]{0.34\linewidth}
\evosechead{EvoGreen}{Constraints}
\begin{evosoftbox}
\scriptsize
\[
0 \le n \le 10^5
\]
\[
0 \le \texttt{original}[i] \le 10^9
\]
\[
0 \le u_i \le v_i \le 10^9
\]
\end{evosoftbox}
\end{minipage}\hfill
\begin{minipage}[t]{0.62\linewidth}
\evosechead{EvoGreen}{Examples}
\begin{evosoftbox}
\scriptsize
\evoex{original=[1,2]; bounds=[[1,2],[2,3]]}{1}
\evoex{original=[1,2]; bounds=[[1,1],[3,3]]}{0}
\evoex{original=[1,2,3]; bounds=[[1,10],[1,10],[1,10]]}{6}
\end{evosoftbox}
\end{minipage}
\end{evoproblemcard}

\vspace{0.35em}

\begin{tcolorbox}[
  enhanced,
  width=\textwidth,
  colback=EvoLightGray,
  colframe=EvoBorderGray,
  boxrule=0.55pt,
  arc=6pt,
  left=8pt,right=8pt,top=5pt,bottom=5pt
]
\footnotesize
\RaggedRight
\textbf{Mutation and algorithmic lift.}
The seed problem has a single additive offset, so it is solved by interval intersection in $O(N)$ time.
The evolved variant changes the relation to XOR, yielding constraints of the form
$u_i \le x\oplus p_i \le v_i$.
Because XOR ranges are not contiguous, simple interval intersection fails, and a bitwise digit-DP or trie-style method is needed.
\end{tcolorbox}
\label{fig:lcb-example-1}

\FloatBarrier

\small
\RaggedRight
\setlength{\emergencystretch}{2em}


\begin{evoproblemcard}
\evocardheader{EvoBlue}{LCB Seed: Uniqueness}{whole array $\cdot$ O(N)}

\begin{minipage}[t]{0.58\linewidth}
\evosechead{EvoBlue}{Task}
Given $N$ people with values $A_i$, find the label of the person whose value is
unique among all people and is numerically largest. If no value is unique,
output $-1$.

\evosechead{EvoBlue}{Core Structure}
Count global frequencies, then select the maximum value with frequency one and
return its index.
\end{minipage}
\hfill
\begin{minipage}[t]{0.38\linewidth}
\evosechead{EvoBlue}{Constraints}
\begin{evosoftbox}
\scriptsize
\[
1 \le N \le 3\times 10^5,
\qquad
1 \le A_i \le 10^9.
\]
\end{evosoftbox}

\evosechead{EvoBlue}{Examples}
\begin{evosoftbox}
\scriptsize
\evoex{A=[2,9,9,7,9,2,4,5,8]}{9}
\evoex{A=[1e9,1e9,998244353,998244353]}{-1}
\end{evosoftbox}
\end{minipage}
\end{evoproblemcard}

\begin{evoproblemcard}
\evocardheader{EvoGreen}{Evolved \#1: Maximum Unique Value Query}{subarray queries}

\begin{minipage}[t]{0.58\linewidth}
\evosechead{EvoGreen}{Task}
Given an array $A$ and $Q$ range queries $(L,R)$, output the index of the
largest value that appears exactly once in $A[L..R]$. If no such value exists,
output $-1$.

\evosechead{EvoGreen}{Algorithmic Lift}
The uniqueness test becomes range-dependent: each query has its own frequency
profile, requiring offline or data-structural reasoning rather than one global
pass.
\end{minipage}
\hfill
\begin{minipage}[t]{0.38\linewidth}
\evosechead{EvoGreen}{Constraints}
\begin{evosoftbox}
\scriptsize
\[
1 \le N,Q \le 5\times 10^5,
1 \le A_i \le 10^9.
\]
\end{evosoftbox}

\evosechead{EvoGreen}{Examples}
\begin{evosoftbox}
\scriptsize
\evoexample{A=[10,10,20,20]; q=(1,4)}{-1}
\evoexample{A=[1,2,3,2,1]; q=(1,5),(1,2)}{3,2}
\evoexample{A=[2,9,9,7,9,2,4,5,8]; q=(1,9)}{9}
\end{evosoftbox}
\end{minipage}
\end{evoproblemcard}

\begin{evoproblemcard}
\evocardheader{EvoPurple}{Evolved \#2: Maximum Unique Path Value}{tree paths}

\begin{minipage}[t]{0.58\linewidth}
\evosechead{EvoPurple}{Task}
Given a tree with node values $W_i$, each query gives nodes $u,v$. On the path
between them, output the node index whose value is largest among values
appearing exactly once on that path. If none exists, output $-1$.

\evosechead{EvoPurple}{Algorithmic Lift}
The range-query setting is lifted from arrays to tree paths, so queries depend
on path decomposition and tree topology rather than contiguous intervals.
\end{minipage}
\hfill
\begin{minipage}[t]{0.38\linewidth}
\evosechead{EvoPurple}{Constraints}
\begin{evosoftbox}
\scriptsize
\[
1 \le N,Q \le 10^5,
\qquad
0 \le W_i \le 10^9.
\]
The input graph is a tree.
\end{evosoftbox}

\evosechead{EvoPurple}{Examples}
\begin{evosoftbox}
\scriptsize
\evoexample{edge 1--2; W=[10,10]; q=(1,2)}{-1}
\evoexample{chain 1--2--3; W=[10,20,10]; q=(1,3)}{2}
\evoexample{chain; W=[1,2,3,2,1];\\ q=(1,5),(1,2)}{3,2}
\end{evosoftbox}
\end{minipage}
\end{evoproblemcard}

\vspace{-1.5cm}
\begin{tcolorbox}[
  enhanced,
  colback=EvoLightGray,
  colframe=EvoBorderGray,
  boxrule=0.55pt,
  arc=6pt,
  left=8pt,right=8pt,top=5pt,bottom=5pt
]
\footnotesize
\RaggedRight
\textbf{Suite-level mutation pattern.}
The suite preserves the semantic target: identify the largest value that is
unique in the relevant scope and return its index. The scope is progressively
generalized from the whole array, to a subarray, and finally to a tree path.
\end{tcolorbox}

\label{fig:lcb-example-2}

\FloatBarrier

\subsection{SciCode Examples}

\newtcolorbox{evocodebox}{%
  enhanced,
  colback=EvoLightGray,
  colframe=EvoBorderGray,
  boxrule=0.45pt,
  arc=5pt,
  left=6pt,right=6pt,top=5pt,bottom=5pt,
  before upper={\ttfamily\scriptsize\RaggedRight\setlength{\parindent}{0pt}\setlength{\emergencystretch}{2em}}
}

\begin{evoproblemcard}
\RaggedRight
\evocardheader{EvoBlue}{SciCode Seed: 4th-Order Runge-Kutta Integrator}{forward simulation}

\evosechead{EvoBlue}{Task}
Implement a classical fourth-order Runge-Kutta (RK4) integrator from scratch
for a driven damped pendulum. The routine takes a derivative function, an
initial state, and fixed time-step information, and returns the full
state-space trajectory over $n$ integration steps.

\vspace{0.12em}
\begin{minipage}[t]{0.58\linewidth}
\evosechead{EvoBlue}{Function Interface}
\begin{evocodebox}
def runge\_kutta\_4th\_order(\\
\ \ \ \ f, state, t0, dt, n, g, L, beta, A, alpha\\
):\\
\ \ \ \ """Run an RK4 integrator for pendulum motion."""
\end{evocodebox}

\evosechead{EvoBlue}{Input / Output}
\begin{evosoftbox}
\footnotesize
\textbf{Input:} derivative function \texttt{f}, initial state
\texttt{state = [theta, omega]}, initial time \texttt{t0}, step size
\texttt{dt}, number of steps \texttt{n}, and pendulum parameters
\texttt{g, L, beta, A, alpha}.

\vspace{0.2em}
\textbf{Output:} \texttt{trajectory}, a 2D NumPy array of shape
\texttt{(n+1, 2)} containing the simulated states.
\end{evosoftbox}

\end{minipage}\hfill
\begin{minipage}[t]{0.38\linewidth}
\evosechead{EvoBlue}{Example Test}
\begin{evocodebox}
state0 = np.array([0.1, 0.0])\\
traj = runge\_kutta\_4th\_order(\\
\ \ \ \ pendulum\_derivs, state0,\\
\ \ \ \ 0.0, 0.01, 100,\\
\ \ \ \ 9.81, 1.0, 0.1, 0.0, 0.0\\
)\\
assert traj.shape == (101, 2)
\end{evocodebox}

\end{minipage}
\end{evoproblemcard}

\vspace{0.35em}

\begin{evoproblemcard}
\RaggedRight
\evocardheader{EvoGreen}{Evolved: Fit ODE Trajectory with Damped Gauss-Newton}{inverse problem}

\evosechead{EvoGreen}{Task}
Implement a nonlinear inverse solver that estimates both the unknown initial
state and unknown ODE parameters from sparse observations. The solver must
simulate the ODE trajectory using fixed-step RK4 between observation times and
fit the unknown variables by damped Gauss-Newton with finite-difference
Jacobians and backtracking line search.

\vspace{0.12em}
\begin{minipage}[t]{0.58\linewidth}
\evosechead{EvoGreen}{Function Interface}
\begin{evocodebox}
def fit\_ode\_trajectory\_gauss\_newton(\\
\ \ \ \ f, state\_guess, param\_guess, t\_obs, observations,\\
\ \ \ \ steps\_per\_unit=100, max\_iter=15, tol=1e-10,\\
\ \ \ \ fd\_eps=1e-7, ridge=1e-10\\
):\\
\ \ \ \ """Fit an ODE trajectory by damped Gauss-Newton."""
\end{evocodebox}

\evosechead{EvoGreen}{Input / Output}
\begin{evosoftbox}
\footnotesize
\textbf{Input:} ODE right-hand side \texttt{f}, guesses for the initial
state and parameters, observation times \texttt{t\_obs}, observed trajectory
array \texttt{observations}, and numerical optimization hyperparameters.

\vspace{0.2em}
\textbf{Output:} a tuple containing the fitted initial state, fitted parameter vector, and simulated trajectory evaluated at the
observation times.
\end{evosoftbox}

\end{minipage}\hfill
\begin{minipage}[t]{0.38\linewidth}
\evosechead{EvoGreen}{Example Test}
\begin{evocodebox}
def exp\_growth(x, t, theta):\\
\ \ \ \ return np.array([theta[0] * x[0]])\\[0.3em]
t\_obs = np.array([0.0, 0.2, 0.55, 1.1])\\
obs = np.exp(-0.8 * t\_obs)[:, None] * 2.5\\[0.3em]
x\_hat, theta\_hat, traj =\\
fit\_ode\_trajectory\_gauss\_newton(\\
\ \ \ \ \ \ \ \ exp\_growth, \\
\ \ \ \ \ \ \ \ np.array([1.0]), \\
\ \ \ \ \ \ \ \ np.array([-0.3]),\\
\ \ \ \ \ \ \ \ t\_obs, \\
\ \ \ \ \ \ \ \ obs, \\
\ \ \ \ \ \ \ \ steps\_per\_unit=200\\
\ \ \ \ )\\
assert traj.shape == obs.shape
\end{evocodebox}

\end{minipage}
\end{evoproblemcard}

\vspace{0.35em}

\begin{tcolorbox}[
  enhanced,
  width=\textwidth,
  colback=EvoLightGray,
  colframe=EvoBorderGray,
  boxrule=0.55pt,
  arc=6pt,
  left=8pt,right=8pt,top=5pt,bottom=5pt
]
\footnotesize
\RaggedRight
\textbf{Mutation and algorithmic lift.}
The seed task performs forward simulation for a fixed ODE using the classical
RK4 integrator. The evolved task preserves the ODE-trajectory setting but turns a direct numerical integration task into a
full nonlinear parameter-estimation pipeline.
\end{tcolorbox}

\label{fig:scicode-ode-fit-example}

\begin{evoproblemcard}
\RaggedRight
\evocardheader{EvoBlue}{SciCode Seed: Compute the $n$-Tangle of an Even-Qubit Pure State}{quantum states}

\evosechead{EvoBlue}{Task}
Implement a function that computes the $n$-tangle of an even-numbered
$n$-qubit pure state $\lvert \psi \rangle$.

The input \texttt{psi} is a one-dimensional array of length $2^n$ representing
the state vector in the computational basis. You may assume that $n$ is even.

\vspace{0.12em}
\begin{minipage}[t]{0.58\linewidth}
\evosechead{EvoBlue}{Function Interface}
\begin{evocodebox}
def n\_tangle(psi):\\
\ \ \ \ """\\
\ \ \ \ Compute the n-tangle of an even-n-qubit pure state.\\
\ \ \ \ """
\end{evocodebox}

\evosechead{EvoBlue}{Input / Output}
\begin{evosoftbox}
\footnotesize
\textbf{Input:} \texttt{psi : np.ndarray}, a one-dimensional array of length
$2^n$ representing a pure quantum state.

\vspace{0.2em}
\textbf{Output:} \texttt{tangle : float}, the $n$-tangle of the input state.
\end{evosoftbox}
\end{minipage}\hfill
\begin{minipage}[t]{0.38\linewidth}
\evosechead{EvoBlue}{Example Tests}
\begin{evocodebox}
MaxEnt = np.array([1, 0, 0, 1]) / np.sqrt(2)

assert np.allclose(
    n\_tangle(MaxEnt), 1.0
)

product\_state = np.kron(
    np.array([0, 1]),
    np.array([0.8, 0.6])
)

assert np.allclose(
    n\_tangle(product\_state), 0.0
)
\end{evocodebox}
\end{minipage}
\end{evoproblemcard}

\vspace{0.35em}

\begin{evoproblemcard}
\RaggedRight
\evocardheader{EvoGreen}{Evolved: Maximum Pairwise CHSH Bell Value}{pairwise optimization}

\evosechead{EvoGreen}{Task}
Implement a function that computes the largest CHSH Bell value attainable
by any pair of qubits in an $n$-qubit quantum state.

The input may be either a pure-state vector or a density matrix, and both
real and complex inputs are allowed.

\vspace{0.12em}
\begin{minipage}[t]{0.58\linewidth}
\evosechead{EvoGreen}{Function Interface}
\begin{evocodebox}
def max\_pairwise\_chsh(state):\\
\ \ \ \ """\\
\ \ \ \ Compute the maximum CHSH Bell value over all\\
\ \ \ \ two-qubit subsystems.\\
\ \ \ \ """
\end{evocodebox}

\evosechead{EvoGreen}{Input / Output}
\begin{evosoftbox}
\footnotesize
\textbf{Input:} \texttt{state : np.ndarray}, either a state vector of length
$2^n$ or a density matrix of shape $(2^n, 2^n)$.

\vspace{0.2em}
\textbf{Output:} \texttt{value : float}, the largest CHSH Bell value among
all distinct pairs of qubits.
\end{evosoftbox}
\end{minipage}\hfill
\begin{minipage}[t]{0.38\linewidth}
\evosechead{EvoGreen}{Example Tests}
\begin{evocodebox}
bell = np.array(
    [1, 0, 0, 1], dtype=float
) / np.sqrt(2)

assert np.isclose(
    max\_pairwise\_chsh(bell),
    2 * np.sqrt(2)
)

ghz = np.zeros(8, dtype=float)
ghz[0] = ghz[7] = 1 / np.sqrt(2)

assert np.isclose(
    max\_pairwise\_chsh(ghz), 2.0
)
\end{evocodebox}
\end{minipage}
\end{evoproblemcard}

\vspace{0.35em}

\begin{tcolorbox}[
  enhanced,
  width=\textwidth,
  colback=EvoLightGray,
  colframe=EvoBorderGray,
  boxrule=0.55pt,
  arc=6pt,
  left=8pt,right=8pt,top=5pt,bottom=5pt
]
\footnotesize
\RaggedRight
\textbf{Mutation and algorithmic lift.}
The seed task computes a single global entanglement quantity for an even-qubit
pure state. The evolved task instead optimizes over all two-qubit subsystems,
searching for the largest attainable CHSH Bell value. This shifts the problem
from evaluating one global scalar functional to analyzing many candidate
subsystems and comparing their pairwise nonlocal correlations.
\end{tcolorbox}

\label{fig:scicode-example-1}

\FloatBarrier

\section{Examples of Evolution Trajectory}

\begin{figure}[H]
    \centering
    \includegraphics[width=0.95\linewidth]{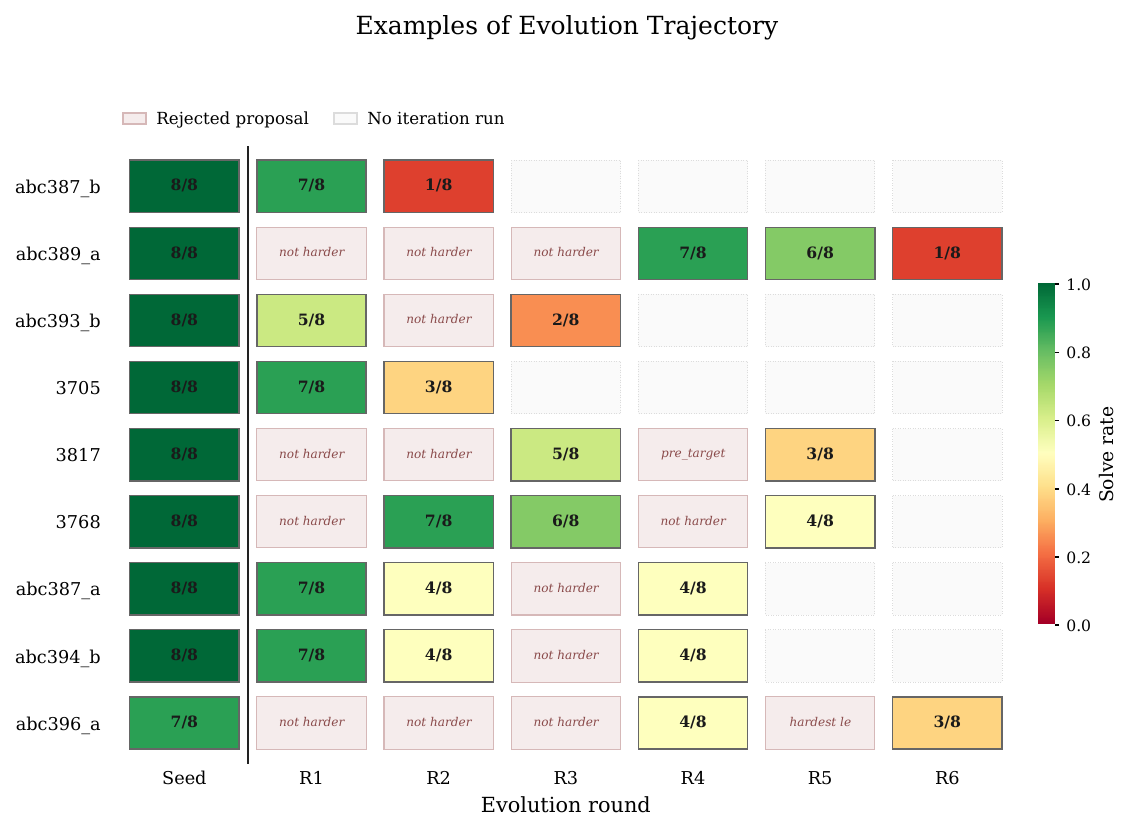}
    \caption{
    \textbf{Example evolution trajectories produced by \methodname.}
    Each row shows one seed problem, labeled by question ID.
    The leftmost column reports the seed solve rate, and subsequent columns
    (\textsf{R1}, \textsf{R2}, \ldots) show accepted evolution rounds in
    chronological order. Each cell reports
    \(\text{passes}/\text{attempts}\), pooled across all target models, where
    \(\text{attempts}\) equals the number of target models times
    \texttt{target\_eval\_k}; cell color encodes solve rate, with red
    indicating harder problems and green indicating easier ones. The dark path
    traces the accepted lineage within each row, highlighting the monotonic
    decrease in solve rate, i.e., the monotonic increase in empirical
    difficulty. Faint pink cells denote rejected proposals: \emph{not harder}
    means the candidate did not exceed the required difficulty level, and
    \emph{judge} means it failed final LLM-judge review. Light gray cells
    indicate that no further iteration was run.
    Overall, the trajectories show that \methodname moves near-saturated seeds
    into a useful difficulty regime, while the rejection mechanism filters
    candidates that are too easy or invalid.
    }
    \label{fig:appx_evolution_trajectory}
\end{figure}

In this appendix section, we append example trajectories of Gemini-3-Flash on LiveCodeBench Easy split, shown in Figure~\ref{fig:appx_evolution_trajectory}.

\section{Prompt Templates for \methodname}
In this section, we provide representative prompt templates used by the main generation components in \methodname. These include the solution mutator, which proposes a harder reference solution; the statement writer, which converts the solution into a complete problem statement; and the test generator, which produces validated tiered test inputs for evaluation.

\newtcblisting{promptbox}[2][]{ enhanced, breakable, colback=PromptBack, colframe=PromptBlue, coltitle=white, fonttitle=\bfseries, title={#2}, boxrule=0.8pt, arc=1.5mm, left=1.5mm, right=1.5mm, top=1mm, bottom=1mm, listing only, listing options={ basicstyle=\ttfamily\tiny, breaklines=true, breakatwhitespace=false, columns=fullflexible, keepspaces=true, showstringspaces=false, tabsize=2 }, #1 }

\begin{promptbox}{Solution Mutator}
You are an expert competitive programmer and benchmark designer.
Your job is to mutate an existing solution into a structurally different,
more complex solution code for a harder problem. Return ONLY valid JSON.

You are mutating a seed problem into a harder variant.

=== PARENT ===
Problem statement:
{parent_problem}

Solution code:
{parent_solution}

=== DIFFICULTY CONTEXT ===
Seed difficulty level: {seed_difficulty_level}

Parent difficulty level: {parent_difficulty_level}

Best level reached for this seed:
{best_level_reached}

Your goal:
reach level {target_level_goal} or higher

=== EVOLUTION MEMORY ===
What worked, what failed, and why:
{memory_context}

=== ALREADY ACCEPTED ACROSS ALL SEEDS ===
Accepted ideas from this run. Propose a dominant algorithmic idea that is
clearly distinct from every entry below.

{global_accepted_context}

=== TARGET MODEL WEAKNESSES ===
Reasoning errors to exploit:
{target_weaknesses}

=== TASK ===
Produce a NEW problem whose difficulty lift is real, natural, genuine, and
algorithmic.

The parent information may be shown in MINIMAL-SPEC mode. If so, treat the
parent solution code as the primary source of truth and use the compact semantic
scaffold only to recover intended semantics and I/O. Do not stay close to the
original wording or story.

Your goal is to move the problem from difficulty level {parent_difficulty_level}
toward level {target_level_goal}. Use the evolution memory above to guide your
mutation strategy.

Keep free mutation mode, but use that freedom to search for a SINGLE DOMINANT
algorithmic lift.

=== Additional Guidance ===
{Examples of valid and invalid mutations}
{Output constraints}

\end{promptbox}

\begin{promptbox}
{Problem Statement Writer}
You are a competitive programming problem setter.
Given a solution and sample input/output pairs, write a clear, natural problem statement.
Return ONLY valid JSON.

Solution code:
{solution_code}

Mutation notes:
{mutation_notes}

Algorithm used: {approach}

Sample input/output pairs:
{sample_io}

Write a competitive-programming-style problem statement for this solution.
Requirements:
- The FINAL solution code is the primary source of truth. Use mutation notes only as hints; if they conflict with the code, ignore the notes and follow the code.
- Natural, unambiguous prose (as in LeetCode/Codeforces/AtCoder)
- Include: problem description, input format, output format, constraints, and sample input/output (embed the provided sample I/O pairs as "Sample Input / Sample Output" sections in the statement body)
- Do NOT hint at the algorithm or data structure used
- If sample I/O pairs are provided, the statement must be consistent with ALL of them
- If no sample I/O is provided, infer a valid and unambiguous input/output spec from the solution code
- Answer must be uniquely determined by the input. If the underlying task admits multiple equally valid answers under different deterministic strategies (e.g., merging equal adjacent values, choosing among tied minima/maxima, selecting any valid partition, returning any optimal subset), you MUST spell out the exact tie-breaking rule the solution uses (e.g., "choose the leftmost such pair", "return the smallest index", "output the lexicographically smallest sequence"). Infer the tie-breaking rule by reading the reference solution code. Do not guess and do not paper over the ambiguity.
- Do NOT write blanket phrases like "any one may be chosen", "the answer is uniquely determined regardless of tie-breaking", or "any valid answer is accepted". If the answer truly does not depend on tie-breaking, prove it by stating the concrete invariant (e.g., "the sum is invariant under order of operations because addition is associative"); if you cannot state such an invariant, treat the task as ambiguous and add an explicit tie-breaking rule.
- Every input variable must have explicit bounds in the constraints section appropriate to its type (e.g. integer range, string length, number of elements). Read them from the solution code if unsure
- Use plain text for all math
- Do NOT use any LaTeX or math markup of any kind
- For every operation type, copy the exact token order from the code's parser. Never swap argument order between the description, input format, and samples.
- Use standard line-based competitive-programming formatting only. Do not invent dual formats, optional parsing conventions, or ambiguous line structure.
- Do not claim stronger constraints than the code supports. If the code is only correct for a simpler/static version, do not write a dynamic/online statement.
- If unsure, use fewer public examples rather than inventing a risky one.
\end{promptbox}

\begin{promptbox}
{Test Generator}
You are a senior competitive-programming test-data engineer.
You write standalone Python 3 programs that generate deterministic, valid test inputs for a given problem.

Your validate() function is the SOLE source of truth for input legality, and the caller performs no secondary format check. Any input you emit that passes validate() will be fed straight to the reference solution and the reference's stdout will be frozen as ground truth. A too-permissive validate() therefore silently corrupts the benchmark; a too-strict validate() starves the suite of inputs.

Your programs must:
  - use a caller-controlled random seed so the same CLI args always produce the same inputs,
  - enforce every declared constraint via validate() exactly as implied by the problem statement AND the reference solution,
  - partition generation into four explicit tiers defined BEHAVIORALLY (see user message): small / medium / large / stress,
  - retry internally on validation failure before giving up on an input,
  - emit a single JSON object on stdout and nothing else.

Return ONLY valid JSON wrapping the program.

Problem statement:
{problem_statement}

Reference solution (authoritative for input format, constraints, and semantics):
{solution_code}

Sample input/output pairs (use these to infer the canonical format):
{sample_io}

Your task: write a SINGLE self-contained Python 3 program that generates test inputs for this problem.

{Tier Definitions}

\end{promptbox}

\section{Reproducibility and Asset Licenses}
\label{app:reproducibility}

\paragraph{Code and configurations.}
We will release the code used to run our evolution pipeline, including the configuration files for LiveCodeBench and SciCode experiments. The released repository will include scripts for seed selection, candidate generation, validation, target-model evaluation, and result aggregation. We will also provide the default hyperparameters used in our experiments, including the target model panels, number of solver attempts, repair budgets, validation thresholds, test-generation settings, and stopping criteria.

\paragraph{Benchmarks and datasets.}
Our experiments use existing public benchmark assets, including LiveCodeBench and SciCode. We cite the original benchmark papers and repositories in the main text. For LiveCodeBench, we use the Version~6 split and report results by seed difficulty. For SciCode, we use the validation split and select the subset of subproblems described in Section~\ref{sec:experiments}. We use these assets only for research evaluation and follow their corresponding licenses and terms of use.

\paragraph{Licenses.}
We use LiveCodeBench under its MIT License and SciCode under the Apache License 2.0. We cite the original benchmark papers and repositories and follow the corresponding licenses and terms of use. For LiveCodeBench, we use the public benchmark data linked from the official repository; for SciCode, we use the Hugging Face dataset released under Apache-2.0. Any released generated tasks, tests, logs, and metadata from our work will include appropriate attribution and license information compatible with the underlying assets.


\end{document}